\documentclass[useAMS,usenatbib]{mn2e}

\usepackage{widetext}

\newcommand{\bea}{\begin{eqnarray}}
\newcommand{\eea}{\end{eqnarray}}

\title[Fully nonlinear and exact perturbations of the Friedmann world model]{Fully nonlinear and exact perturbations of the Friedmann world model}
\author[Jai-chan Hwang and Hyerim Noh]{Jai-chan Hwang$^{1}$\thanks{E-mail:
jchan@knu.ac.kr (JH); hr@kasi.re.kr (HN)} and Hyerim Noh$^{2}$\footnotemark[1]\\
$^{1}$Department of Astronomy and Atmospheric Sciences,
                 Kyungpook National University, Daegu 702-701, Republic of Korea\\
$^{2}$Korea Astronomy and Space Science Institute,
                 Daejeon 305-348, Republic of Korea}
\begin{document}

\date{}

\pagerange{\pageref{firstpage}--\pageref{lastpage}} \pubyear{2002}

\maketitle

\label{firstpage}

\begin{abstract}
In 1988 Bardeen suggested a pragmatic formulation of cosmological
perturbation theory which is powerful in practice to employ various
fundamental gauge conditions easily depending on the character of
the problem. The perturbation equations are presented without fixing
the temporal gauge condition and are arranged so that one can easily
impose fundamental gauge conditions by simply setting one of the
perturbation variables in the equations equal to zero. In this way
one can use the gauge degrees of freedom as an advantage in handling
problems. Except for the synchronous gauge condition, all the other
fundamental gauge conditions completely fix the gauge mode, and
consequently, each variable in such a gauge has a unique gauge
invariant counterpart, so that we can identify the variable as the
gauge-invariant one. Here, we extend Bardeen's linear formulation to
the fully nonlinear order in perturbations, with the gauge advantage
kept intact. The perturbation equations are exact, and from these we can
easily derive the higher order perturbation equations in a gauge-ready form.
We consider scalar- and vector-type perturbations of an ideal fluid
in a flat background; we also present the minimally coupled scalar field and  the multiple components of ideal fluid cases. As applications we present fully nonlinear density
and velocity perturbation equations in Einstein's gravity in the
zero-pressure medium, vorticity generation from pure scalar-type
perturbation, and fluid formulation of a minimally coupled scalar
field, all in the comoving gauge. We also present the equation of
gravitational waves  generated from pure scalar- and vector-type
perturbations.
\end{abstract}


\section{Introduction}
                                             \label{sec:Introduction}

The Friedmann world model, based on assuming spatial homogeneity and
isotropy in Einstein's gravity, is widely accepted as a successful
cosmological model, enduring 90 years of observational and
theoretical advances after Friedmann's original proposition in 1922.
Main observational and theoretical advances have been made in the
angular anisotropies of cosmic microwave background radiation and in
the position and motion of large scale galaxy distribution.
Relativistic perturbation theory is important in providing crucial
testing grounds for matching the theoretical predictions with the
observation, and for the theoretical explanation of the observed
phenomena. The relativistic linear perturbation theory and the
Newtonian exact and nonlinear perturbation theory are generally
accepted to be successful in the current paradigm of modern physical
cosmology. This work is concerned with relativistic fully nonlinear
and exact perturbation formulation in the Friedmann world model.

The cosmological linear perturbation theory in the Friedmann's world
model was pioneered by Lifshitz (1946). Lifshitz's analysis was
made in a certain gauge (coordinate) condition known as the
synchronous gauge (Landau \& Lifshitz 1975). A disadvantage of the
synchronous gauge condition was that even after imposing the gauge
condition there remains remnant gauge (coordinate) mode which needs
to be traced carefully. In this way the algebras become
unnecessarily complicated. Often, removing the remnant gauge mode in
the synchronous gauge merely corresponds to taking just another
gauge condition; for example, setting the velocity component of the
pressureless matter (being a gauge-mode in the synchronous gauge)
equal to zero is equivalent to simply taking the velocity component
of the pressureless matter equal to zero (the comoving gauge of the
pressureless matter); however, as will become clear in this work,
even such a cure in the synchronous gauge is available only to the
linear order in the presence of a zero-pressure fluid (Hwang \& Noh
2006). The remnant gauge mode in the synchronous gauge causes quite
a troublesome matter to handle in the nonlinear perturbation theory.

Other gauge conditions free from the remnant gauge modes were
proposed by Harrison (1967) for the zero-shear gauge (often known as
the longitudinal or conformal-Newtonian gauge), and Nariai (1969)
for the comoving gauge; see also Hawking (1966), Sachs \& Wolfe
(1967), and Field \& Shepley (1968). There are, in fact, infinitely
many other gauge conditions which are free from such a complication:
i.e., free from the remnant gauge mode, and thus equivalently gauge
invariant, see later. Systematic introduction of several different
gauge conditions with explicit display of gauge-invariant
combination of variables was made by Bardeen (1980) with a huge
success in later applications in the literature; see Kodama \&
Sasaki (1984) for a review.

In a less known work, in 1988 Bardeen suggested a pragmatic way of
deploying the gauge conditions depending on the problems. As in
other gauge theories the gauge choice is the degree of freedom which
can be employed depending on the advantages in achieving either
mathematical simplification or plausible physical interpretation.
Bardeen has arranged equations so that the fundamental gauge
conditions can be implemented easily. Bardeen's formulation of
linear perturbation theory was extended in Hwang (1991), Hwang \&
Noh (2001, 2005), and to the second order perturbations in Noh \&
Hwang (2004) and Hwang \& Noh (2007).

Our aim in this work is to extend the Bardeen's formulation to the
exact and fully nonlinear order in perturbations keeping the gauge
advantages intact. We will display some applications, but since the
main point is to present the new and powerful nonlinear perturbation
equations, we will show detailed steps needed for the derivation in
the Appendices \ref{sec:ADM} and \ref{sec:derivation}. The main equations are presented in Section \ref{sec:equations}, and in the Appendices \ref{sec:Multiple} and \ref{sec:MSF} for multi-component fluid case and the minimally coupled scalar field case, respectively.

In Section \ref{sec:convention} we review Bardeen's formulation and
the gauge strategy in nonlinear perturbations. In Section
\ref{sec:equations} the exact and fully nonlinear equations are
presented {\it assuming} scalar- and vector-type perturbations of an
ideal fluid in a flat background, but without fixing the temporal
gauge condition. In Section \ref{sec:3rd-order} we present equations
valid to the third order in perturbations still without fixing the
temporal gauge condition. In Section \ref{sec:CG} we make several
applications available in the comoving gauge including comparison
with the Newtonian results. In Section \ref{sec:other-gauges} we
consider the cases in other fundamental gauge conditions. In Section
\ref{sec:rotation} we consider vorticity generation from pure
scalar-type perturbation in the comoving gauge. In Section \ref{sec:GW} we present the equation of
gravitational waves generated from pure scalar- and vector type
perturbations which depends on the gauge choice. In Section
\ref{sec:MSF-CG} we analyze the case of a minimally coupled scalar
field in the comoving gauge. We show that in the comoving gauge the ideal fluid equations remain valid for the scalar field with a particularly simple equation of state.
In Section \ref{sec:Discussion} we comment on the possible
future extension of this work.
Appendix \ref{sec:ADM} is a review of the ADM formulation (Arnowitt, et al 1962). In the Appendix \ref{sec:derivation} we present detailed steps needed for deriving the nonlinear equations in exact form. In the Appendix \ref{sec:cov} we present the analysis based on the covariant formulation. In the Appendix \ref{sec:velocities} we introduce and clarify different definitions of the fluid three-velocities. Appendices \ref{sec:Multiple} and \ref{sec:MSF} present the cases of multi-component fluids and the minimally coupled scalar field, respectively. We set $c \equiv 1$ except for Section \ref{sec:equations}.

\section{Convention and gauge strategy}
                                             \label{sec:convention}

Here are our metric and the energy-momentum tensor conventions. We
consider scalar- and vector-type perturbations in a {\it flat}
Robertson-Walker background. The metric can be written as \bea
   & &
       ds^2 = - a^2 \left( 1 + 2 \alpha \right) d \eta^2
       - 2 a^2 \left( \beta_{,i} + B^{(v)}_i \right) d \eta d x^i
   \nonumber \\
   & & \qquad
       + a^2 \left[ \left( 1 + 2 \varphi \right) \delta_{ij}
       + 2 \gamma_{,ij}
       + C^{(v)}_{i,j} + C^{(v)}_{j,i} + 2 C^{(t)}_{ij} \right]
   \nonumber \\
   & & \qquad
       \times
       d x^i d x^j,
   \label{metric}
\eea where $a(\eta)$ is the cosmic scale factor, and we assume
$B^{(v)i}_{\;\;\;\;\;\;,i} \equiv 0 \equiv
C^{(v)i}_{\;\;\;\;\;\;,i}$ (transverse), and
$C^{(t)j}_{\;\;\;\;\;i,j} = 0 = C^{(t)i}_{\;\;\;\;\;i}$
(transverse-tracefree) with indices of $B^{(v)}_i$, $C^{(v)}_i$ and
$C^{(t)}_{ij}$ raised and lowered by $\delta_{ij}$ as the metric; indices $(v)$
and $(t)$ indicate the vector- and tensor-type perturbations,
respectively. Indices $a, b, \dots$ indicate the spacetime indices,
and $i, j, \dots$ indicate the spatial ones; we follow the
convention of Hawking \& Ellis (1973).

The energy momentum tensor is given as (Ehlers 1993; Ellis 1971, 1973)
\bea
   & & \widetilde T_{ab} = \widetilde \mu \widetilde u_a \widetilde u_b
       + \widetilde p \left( \widetilde u_a \widetilde u_b + \widetilde g_{ab} \right)
       + \widetilde q_a \widetilde u_b
       + \widetilde q_b \widetilde u_a
       + \widetilde \pi_{ab},
   \label{Tab-general}
\eea where $\widetilde \mu$ and $\widetilde p$ are the energy
density and the pressure, respectively, $\widetilde  u_a$ is a
normalized fluid four-vector with $\widetilde u^a \widetilde u_a \equiv
-1$, $\widetilde q_a$ is the energy flux with $\widetilde q^a \widetilde u_a \equiv 0$, and $\widetilde \pi_{ab}$ is the anisotropic stress with
$\widetilde \pi_{ab} = \widetilde \pi_{ba}$, $\widetilde \pi^a_a
\equiv 0$, and $\widetilde \pi_{ab} \widetilde u^b \equiv 0$; tildes
indicate the covariant quantities.
The fluid quantities can be read from the energy-momentum tensor as \bea
   & & \widetilde \mu = \widetilde T_{ab} \widetilde u^a \widetilde u^b, \quad
       \widetilde p = {1 \over 3} \widetilde T_{ab} \widetilde h^{ab},
   \nonumber \\
   & &
       \widetilde q_a = - \widetilde T_{cd} \widetilde u^c \widetilde h^d_a, \quad
       \widetilde \pi_{ab} = \widetilde T_{cd} \widetilde h^c_a \widetilde h^d_b - \widetilde p \widetilde h_{ab}.
   \label{Tab-decomposition}
\eea

In this work we consider an ideal
fluid with $\widetilde \pi_{ab} = 0$, and {\it take} the energy frame by setting $\widetilde q_a \equiv 0$. The energy-momentum tensor becomes
\bea
   & & \widetilde T_{ab} = \widetilde \mu \widetilde u_a \widetilde u_b
       + \widetilde p \left( \widetilde u_a \widetilde u_b + \widetilde g_{ab} \right).
   \label{Tab}
\eea
We set \bea
   & & \widetilde \mu \equiv \mu + \delta \mu
       \equiv \mu \left( 1 + \delta \right), \quad
       \widetilde p \equiv p + \delta p, \quad
       \widetilde u_i \equiv a v_{i},
   \label{fluids}
\eea where $\mu$ and $p$ are the background energy density and
pressure, respectively. We set \bea
   & & v_i \equiv - v_{,i} + v^{(v)}_i,
   \label{v-decomposition}
\eea with $v^{(v)i}_{\;\;\;\;\;\;,i} \equiv 0$; $v_i$ and
$v^{(v)}_i$ are raised and lowered by $\delta_{ij}$ as the metric. More physically motivated definitions of the fluid three-velocities will be presented in the Appendix \ref{sec:velocities}.

In this work, unless mentioned explicitly, we do {\it not} assume the perturbed metric and fluid quantities are small.

The decomposition of an arbitrary spatial vector into longitudinal
and transverse parts as $B_i = \beta_{,i} + B^{(v)}_i$, and a
symmetric spatial tensor into longitudinal, trace, transverse, and
tracefree-transverse parts as $C_{ij} = \varphi \delta_{ij} +
\gamma_{,ij} + {1 \over 2} (C^{(v)}_{i,j} + C^{(v)}_{j,i} ) +
C^{(t)}_{ij}$ are possible (York 1973); here, all spatial indices
are raised and lowered by $\delta_{ij}$ as the metric. We {\it assign} the transverse
part as the vector-type perturbation, and the tracefree-transverse
part as the tensor-type perturbation, and the remaining longitudinal
and trace parts as the scalar-type perturbation. We can show that
the decomposition is possible order by order in perturbation to
nonlinear order. However, only to the linear order in the spatially
homogeneous-isotropic background the three types of perturbations
decouple from each other. To the nonlinear order we have couplings
among the scalar-, vector- and tensor-types of perturbations in the equation level.

Here we {\it ignore} the tensor-type perturbation. Restriction of
our attention only to the scalar- and vector-type perturbations can
be regarded as our main {\it assumption} in this work. In Section
\ref{sec:GW}, though, we will consider the generation of the linear order
tensor-type perturbation from the nonlinear scalar- and vector-type perturbations.

In considering the linear perturbation theory, Bardeen has suggested
to take the spatial gauge condition \bea
   & & \gamma \equiv 0 \equiv C^{(v)}_i,
   \label{spatial-gauge}
\eea but has saved the temporal gauge condition for later use. To
the linear order \bea
   & & \chi \equiv a \beta + a^2 \dot \gamma, \quad
       \Psi^{(v)}_i \equiv B^{(v)}_i + a \dot C^{(v)}_i,
   \label{spatial-GI}
\eea are spatially gauge-invariant, see below equation (\ref{GT}); an
overdot is a time derivative based on background cosmic time $t$
($dt \equiv a d \eta$). Under the spatial gauge condition in
equation (\ref{spatial-gauge}) our metric convention becomes \bea
   & &
       ds^2 = - a^2 \left( 1 + 2 \alpha \right) d \eta^2
       - 2 a \chi_{i} d \eta d x^i
   \nonumber \\
   & & \qquad
       + a^2 \left( 1 + 2 \varphi \right) \delta_{ij} d x^i d x^j,
   \label{metric-convention}
\eea where we set \bea
   & & \chi_i \equiv \chi_{,i} + a \Psi^{(v)}_i
       = a \left( \beta_{,i} + B^{(v)}_i \right).
\eea We may also write $\chi^{(v)}_i \equiv a \Psi^{(v)}_i$. We will
take the metric and the energy-momentum tensor conventions in
equations (\ref{metric-convention}) and (\ref{Tab}) even in
nonlinear perturbation theory. Justification for taking the spatial
gauge condition in equation (\ref{spatial-gauge}) to the nonlinear
order will be made later in this section. It is essentially these
spatial gauge conditions together with ignoring the tensor-type
perturbation which allow our fully nonlinear and exact formulation
available. As will be explained, we do not lose any generality or
convenience by taking these spatial gauge conditions; the only other
alternative choice of spatial gauge condition leaves remnant gauge
modes even from the linear order, see below equation (\ref{GT}).

The scalar-type perturbation equations without taking the temporal
gauge condition are arranged in the following form (Bardeen 1988)
\bea
   & & \kappa \equiv 3 H \alpha - 3 \dot \varphi - {\Delta \over a^2} \chi,
   \label{eq1-linear} \\
   & & 4 \pi G \delta \mu + H \kappa + {\Delta \over a^2} \varphi
       = 0,
   \label{eq2-linear} \\
   & & \kappa
       + {\Delta \over a^2} \chi - 12 \pi G (\mu + p) a v
       = 0,
   \label{eq3-linear} \\
   & & \dot \kappa + 2 H \kappa
       - 4 \pi G \left( \delta \mu + 3 \delta p \right)
       + \left( 3 \dot H + {\Delta \over a^2} \right) \alpha
       = 0,
   \label{eq4-linear} \\
   & & \dot \chi + H \chi - \varphi - \alpha
       = 0,
   \label{eq5-linear} \\
   & & \delta \dot \mu + 3 H \left( \delta \mu + \delta p \right)
       - \left( \mu + p \right) \left( \kappa - 3 H \alpha
       + {1 \over a} \Delta v \right)
   \nonumber \\
   & & \qquad
       = 0,
   \label{eq6-linear} \\
   & & {[a^4 (\mu + p) v]^{\displaystyle\cdot} \over a^4(\mu + p)}
       - {1 \over a} \alpha
       - {\delta p \over a (\mu + p)}
       = 0,
   \label{eq7-linear}
\eea where $H \equiv \dot a/a$; here we have {\it assumed} the flat
background and an ideal fluid. These equations are arranged {\it
without} taking the temporal gauge condition. One major advantage of
this arrangement is that the equations are designed so that we can
readily impose the various fundamental gauge condition by simply setting
one of the perturbation variables equal to zero. The vector-type perturbation
equations are (Bardeen 1980) \bea
   & & {\Delta \over 2 a^2} \Psi^{(v)}_i
       + 8 \pi G \left( \mu + p \right) v^{(v)}_i = 0,
   \label{eq3-vector} \\
   & & \left[ a^4 \left( \mu + p \right) v^{(v)}_i \right]^{\displaystyle\cdot}
       = 0.
   \label{eq7-vector}
\eea These vector-type equations are gauge-invariant. In this work
we will present the exact and fully nonlinear extension of these
equations: see equations (\ref{eq1})-(\ref{eq7}).

The spatial gauge (congruence) condition in equation
(\ref{spatial-gauge}) is a {\it unique} choice without remnant
spatial gauge mode after taking the gauge condition; see below equation (\ref{GT}). Thus, we do not
lose any advantage upon our choice of the spatial gauge condition.
On such a spatial gauge condition, the remaining variables can be
regarded as spatially gauge-invariant ones (Bardeen 1988); see
equation (\ref{spatial-GI}).

Concerning the temporal gauge condition, however, we have several
fundamental gauge conditions available most of which remove the
temporal gauge mode completely. We have the following fundamental
gauge conditions \bea
   & & {\rm comoving \; gauge:}              \hskip 2.0cm     v \equiv 0,
   \nonumber \\
   & & {\rm zero\!-\!shear \; gauge:}        \hskip 1.72cm     \chi \equiv 0,
   \nonumber \\
   & & {\rm uniform\!-\!curvature \; gauge:} \hskip .56cm      \varphi \equiv 0,
   \nonumber \\
   & & {\rm uniform\!-\!expansion \; gauge:} \hskip .52cm      \kappa \equiv 0,
   \nonumber \\
   & & {\rm uniform\!-\!density \; gauge:}   \hskip .94cm      \delta \equiv 0,
   \nonumber \\
   & & {\rm synchronous \; gauge:}           \hskip 1.58cm     \alpha \equiv 0.
   \label{temporal-gauges}
\eea Also available as the gauge conditions are any
non-gauge-invariant combination of these gauge conditions, thus we
could have infinite number of different temporal gauge (spatial
hyperspace or slicing) conditions available. As a consequence, we
can manage an arbitrary form of differential equation for any
non-gauge-invariant variable using a certain, perhaps {\it ad hoc},
choice of the gauge condition (Hwang et al. 2010).

Bardeen's arrangement of equations apparently allows the simple
adaptation of any fundamental gauge conditions: we simply set a
perturbation variable equal to be zero. Except for the synchronous
gauge the other temporal gauge conditions together with the spatial
gauge condition in equation (\ref{spatial-gauge}) completely remove
the gauge (coordinate transformation) degrees of freedom. Thus, each
of the remaining perturbation variables has a unique counterpart of
gauge-invariant combination involving the variable concerned and the
variables used in the spatial and temporal gauge conditions; see
below equation (\ref{GT}). Therefore, all the variables in such a
gauge condition can be equivalently regarded as the corresponding
gauge-invariant variables.

The gauge conditions in equations (\ref{spatial-gauge}) and
(\ref{temporal-gauges}), and the above statements about the gauge
issue remain valid to the fully nonlinear order as long as we take
the perturbation approach: this was shown in Section VI of
Noh \& Hwang (2004). In the following, we explain it again.

We consider gauge transformation properties under $\widehat x^a =
x^a + \widetilde \xi^a (x^e)$ with $\widetilde \xi^0 = \xi^0$,
$\widetilde \xi^i = \xi^i$, and $\xi_i \equiv \xi_{,i}/a +
\xi^{(v)}_i$ with $\xi^{(v)i}_{\;\;\;\;\;\;,i} \equiv 0$; index of $\xi^i$ is
raised and lowered by $\delta_{ij}$ as the metric. To the linear order we have
(Bardeen 1988; Noh \& Hwang 2004) \bea
   & & \widehat \delta = \delta - {\mu^\prime \over \mu} \xi^0, \quad
       \widehat v = v - \xi^0, \quad
       \widehat \alpha = \alpha - {1 \over a} \left( a \xi^0 \right)^\prime,
   \nonumber \\
   & &
       \widehat \beta = \beta - \xi^0 + \left( {1 \over a} \xi \right)^\prime, \quad
       \widehat \gamma = \gamma - {1 \over a} \xi, \quad
       \widehat \varphi = \varphi - a H \xi^0,
   \nonumber \\
   & &
       \widehat \chi = \chi - a \xi^0, \quad
       \widehat \kappa = \kappa + \left( 3 \dot H
       + {\Delta \over a^2} \right) a \xi^0,
   \nonumber \\
   & &
       \widehat B^{(v)}_i = B^{(v)}_i + \xi^{(v)\prime}_i, \quad
       \widehat C^{(v)}_i = C^{(v)}_i - \xi^{(v)}_i,
   \label{GT}
\eea where a prime denotes the time derivative based on $\eta$.
Apparently $\gamma \equiv 0 \equiv C^{(v)}_i$ in all coordinates
leaves $\xi_i = 0$, thus fixing the spatial gauge degree of freedom
completely; the only other choice taking $\beta \equiv 0 \equiv
B^{(v)}_i$ in all coordinates gives $\xi \neq 0 \neq \xi^{(v)}_i$,
thus leaving remnant spatial gauge modes. Similarly, for the
fundamental temporal gauge condition, for example, $v = 0$ in all coordinates
leaves $\xi^0 = 0$, thus fixing the temporal gauge degree of freedom
completely. The following combinations are gauge invariant \bea
   & &
       \varphi_v \equiv \varphi - a H v \equiv - a H v_\varphi,
       \quad
       \varphi_\chi \equiv \varphi - H \chi \equiv - H \chi_\varphi,
   \nonumber \\
   & &
       \varphi_\delta \equiv \varphi
       + {\delta \mu \over 3 (\mu + p)}, \quad
       \delta_v \equiv \delta - a {\dot \mu \over \mu} v,
   \nonumber \\
   & &
       v_\chi \equiv v - {1 \over a} \chi, \quad
       e \equiv \delta p_{\delta \mu}
       \equiv \delta p - {\dot p \over \dot \mu} \delta \mu,
   \label{GI}
\eea etc. This shows a systematic notation of expressing the various
gauge-invariant combinations. This notation is practically useful to
implement the spirit of Bardeen's formulation employing many gauge
conditions which make all variables gauge invariant (Hwang 1991).
The gauge-invariant combination, for example, $\varphi_v$ is the
same as $\varphi$ in the $v \equiv 0$ hypersurface condition, thus
$\varphi_v = \varphi |_{v \equiv 0}$. The temporal gauge condition,
for example, $v = 0$ fixes the temporal gauge mode completely. Thus,
any perturbation variables in that gauge, for example $\varphi$, can
be equivalently regarded as a temporally gauge invariant ones, i.e.,
$\varphi |_{v \equiv 0} = \varphi_v$. Similar complete gauge fixings are true
for the other fundamental temporal gauge conditions. The synchronous gauge
($\alpha \equiv 0$) is an exception, leaving a remnant temporal
gauge mode $\xi^0 (\eta, {\bf x}) \propto a^{-1}$ even after fixing
the gauge condition.

Now, to the nonlinear order, we may set $\xi^0 = \xi^{0(1)} +
\xi^{0(2)} + \dots$ and $\xi_i = \xi^{(1)}_i + \xi^{(2)}_i + \dots$
where the number inside the parenthesis indicates the perturbation
order. To the second order, the gauge transformation properties of
each variable have the same form as in equations (\ref{GT}) with
additional terms involving quadratic combinations of $\xi^0$, $\xi$
and perturbation variables, all to the linear order. Since each of
the quadratic terms involves $\xi^0$ or $\xi$ to the linear order,
as long as we take the spatial and temporal gauge conditions which
lead to $\xi^{0(1)} = 0 = \xi^{(1)}_i$ (thus the synchronous gauge
is excluded), we have exactly the same gauge transformation
properties in equation (\ref{GT}) now valid for pure second order
variables: for example, we have $\delta \widehat \mu^{(2)} = \delta \mu^{(2)} - \mu_{,0} \xi^{0(2)}$, etc. Thus, by imposing the same (i.e., ones removing
the gauge degrees of freedom completely) gauge conditions now to the
second order, we have  $\xi^{0(2)} = 0 = \xi^{(2)}_i$, thus
leaving any variable in that gauge having a corresponding unique
gauge-invariant counterpart: i.e., $\delta \mu |_{v \equiv 0} = \delta \mu_v$, etc.
Apparently, the same process can be continued to any higher order
perturbations.

Let us elaborate the explanation in previous paragraph using an
example. We consider the gauge transformation property of the energy
density $\widetilde \mu \equiv \widetilde T_{ab} \widetilde u^a
\widetilde u^b$ which is a scalar quantity. Under the gauge
transformation introduced above equation (\ref{GT}), we have
$\widehat {\widetilde \mu} (\widehat x^e) = \widetilde \mu (x^e)$.
Separating the background and perturbation as $\widetilde \mu = \mu
+ \delta \mu$ and comparing $\widehat {\widetilde \mu}$ and
$\widetilde \mu$ at the same spacetime point, say $x^e$, we have
\bea
   & & \delta \widehat \mu (x^e)
       = \delta \mu (x^e)
       - \widetilde \mu_{,a} \widetilde \xi^a
       + \left( {1 \over 2} \widetilde \mu_{,ab} \widetilde \xi^a
       + \widetilde \mu_{,a} \widetilde \xi^a_{\;\; ,b} \right)
       \widetilde \xi^b
   \nonumber \\
   & & \qquad
       - \left[ {1 \over 2}
       \widetilde \mu_{,a} \widetilde \xi^a_{\;\;,bc}
       \widetilde \xi^b
       + \left( \widetilde \mu_{,a} \widetilde \xi^b \right)_{,c}
       \widetilde \xi^a_{\;\; ,b}
       + {1 \over 6} \widetilde \mu_{,abc}
       \widetilde \xi^a \widetilde \xi^b \right] \widetilde \xi^c
   \nonumber \\
   & & \qquad
       + \dots
       .
\eea
To the linear order we have \bea
   & & \delta \widehat \mu (x^e)
       = \delta \mu (x^e)
       - \mu_{,0} \xi^0,
   \label{GT-mu-lin}
\eea where we used $\mu = \mu (x^0)$ and $\widetilde \xi^0 = \xi^0$
introduced above equation (\ref{GT}). As long as we take the
spatial gauge fixing $\gamma \equiv 0 \equiv C^{(v)}_i$ and take any
one of the temporal gauge fixing in the pool of the fundamental
temporal gauge conditions in equation (\ref{temporal-gauges}),
except for the synchronous gauge, we have $\xi^0 = 0 = \xi^i$ to the
linear order.

To the second order, we have \bea
   & & \delta \widehat \mu (x^e)
       = \delta \mu (x^e)
       - \mu_{,0} \xi^0
       - \delta \mu_{,0} \xi^0
       - \delta \mu_{,i} \xi^i
   \nonumber \\
   & & \qquad
       + \left( {1 \over 2} \mu_{,00} \xi^0
       + \mu_{,0} \xi^0_{\;\;,0} \right) \xi^0
       + \mu_{,0} \xi^0_{\;\;,i} \xi^i.
\eea
Now, as we have $\xi^0 = 0 = \xi^i$ to the linear order in our suggested gauge conditions, the above equation simply leads to \bea
   & & \delta \widehat \mu (x^e)
       = \delta \mu (x^e)
       - \mu_{,0} \xi^0,
\eea
which is the same as equation (\ref{GT-mu-lin}), now valid even to the second order. In a decomposed form we have\bea
   & & \delta \widehat \mu^{(2)} (x^e)
       = \delta \mu^{(2)} (x^e)
       - \mu_{,0} \xi^{0(2)}.
\eea Therefore, by imposing the gauge conditions in the suggested
pool, now to the second order, we again have $\xi^{0(2)} = 0 =
\xi^{i(2)}$. This can be continued to the higher order perturbations
as well. Explicit forms of the gauge transformation properties and
gauge-invariant combinations to the second-order perturbation are
presented in Noh \& Hwang (2004) and Hwang, et al (2012). The gauge
transformation property in the nonlinear perturbation theory is also
studied in Bruni et al (1997), Matarrese et al (1998), Sonego \&
Bruni (1998), Malik \& Wands (2009), and Nakamura (2010).

The names of our gauge conditions can be justified to the nonlinear
order by examining the ADM metric, extrinsic-curvature,
intrinsic-curvature, and the fluid variables presented in the
Appendices B and C. For the comoving gauge with $v \equiv 0$,
ignoring the vector-type perturbation, from equation
(\ref{four-vector}) we have $\widetilde u_i = 0$, thus the fluid
four-vector becomes the normal four-vector. For the zero-shear gauge
with $\chi \equiv 0$, ignoring the vector-type perturbation, from
equation (\ref{extrinsic-curvature}) we have $\overline{K}^i_j = 0$,
thus having vanishing shear of the normal flow vectors $\widetilde
n_a$; we have $\widetilde \sigma^{(n)}_{ij} = - \overline{K}_{ij}$,
see equation (\ref{kinematic-quantities-normal}). For the uniform-curvature gauge with $\varphi
\equiv 0$, from equation (\ref{intrinsic-curvature}) we have
$R^{(h)i}_{\;\;\;\;\;\;jk\ell} = 0$, thus having vanishing curvature
of the spatial hypersurface; in the presence of the background
spatial curvature, we have spatially uniform curvature. For the
uniform-expansion gauge with $\kappa \equiv 0$, from equation
(\ref{extrinsic-curvature}) we have $K = - 3 H$, thus having the
trace of extrinsic curvature uniform; we have $\widetilde
\theta^{(n)} \equiv \widetilde n^c_{\;\; ;c} = - K$, see equation (\ref{kinematic-quantities-normal}). For the uniform-density gauge with $\delta \equiv 0$,
from equation (\ref{fluids}) we have $\widetilde \mu = \mu$, thus
the density becomes uniform in the hypersurface.

\begin{widetext}
\section{Exact and fully nonlinear perturbation equations without taking temporal gauge condition}
                                             \label{sec:equations}

Fully nonlinear extension of equations (\ref{eq1-linear})-(\ref{eq7-vector}) will be presented below; we recover $c$ in this section. These equations are the main result of this work. Based on the ADM equations, the derivation of our fundamental equations is unexpectedly simple. In order to help the reader who will attempt the derivation we review the ADM formulation in the Appendix A, and present detailed steps required for the derivation in the Appendix B.

\noindent
Definition of $\kappa$:
\bea
   & & \kappa
       + 3 H \left( {1 \over {\cal N}} - 1 \right)
       + {1 \over {\cal N} (1 + 2 \varphi)}
       \left[ 3 \dot \varphi
       + {c \over a^2} \left( \chi^k_{\;\;,k}
       + {\chi^{k} \varphi_{,k} \over 1 + 2 \varphi} \right)
       \right]
       = 0.
   \label{eq1}
\eea
ADM energy constraint:
\bea
   & & - {3 \over 2} \left( H^2 - {8 \pi G \over 3 c^2} \widetilde \mu
       - {\Lambda c^2 \over 3} \right)
       + H \kappa
       + {c^2 \Delta \varphi \over a^2 (1 + 2 \varphi)^2}
       = {1 \over 6} \kappa^2
       - {4 \pi G \over c^2} \left( \widetilde \mu + \widetilde p \right)
       \left( \widehat \gamma^2 - 1 \right)
       + {3 \over 2} {c^2 \varphi^{,i} \varphi_{,i} \over a^2 (1 + 2 \varphi)^3}
       - {c^2 \over 4} \overline{K}^i_j \overline{K}^j_i.
   \label{eq2}
\eea
ADM momentum constraint:
\bea
   & & {2 \over 3} \kappa_{,i}
       + {c \over 2 a^2 {\cal N} ( 1 + 2 \varphi )}
       \left( \Delta \chi_i
       + {1 \over 3} \chi^k_{\;\;,ik} \right)
       + {8 \pi G \over c^4} \left( \widetilde \mu + \widetilde p \right)
       a \widehat \gamma v_{i}
   \nonumber \\
   & & \qquad
       =
       {c \over a^2 {\cal N} ( 1 + 2 \varphi)}
       \Bigg\{
       \left( {{\cal N}_{,j} \over {\cal N}}
       - {\varphi_{,j} \over 1 + 2 \varphi} \right)
       \left[ {1 \over 2} \left( \chi^{j}_{\;\;,i} + \chi_i^{\;,j} \right)
       - {1 \over 3} \delta^j_i \chi^k_{\;\;,k} \right]
   \nonumber \\
   & & \qquad
       - {\varphi^{,j} \over (1 + 2 \varphi)^2}
       \left( \chi_{i} \varphi_{,j}
       + {1 \over 3} \chi_{j} \varphi_{,i} \right)
       + {{\cal N} \over 1 + 2 \varphi} \nabla_j
       \left[ {1 \over {\cal N}} \left(
       \chi^{j} \varphi_{,i}
       + \chi_{i} \varphi^{,j}
       - {2 \over 3} \delta^j_i \chi^{k} \varphi_{,k} \right) \right]
       \Bigg\}.
   \label{eq3}
\eea
Trace of ADM propagation:
\bea
   & & - 3 {1 \over {\cal N}} \dot H
       - 3 H^2
       - {4 \pi G \over c^2} \left( \widetilde \mu + 3 \widetilde p \right)
       + \Lambda c^2
       + {1 \over {\cal {\cal N}}} \dot \kappa
       + 2 H \kappa
       + {c^2 \Delta {\cal N} \over a^2 {\cal N} (1 + 2 \varphi)}
   \nonumber \\
   & & \qquad
       = {1 \over 3} \kappa^2
       + {8 \pi G \over c^2} \left( \widetilde \mu + \widetilde p \right)
       \left( \widehat \gamma^2 - 1 \right)
       - {c \over a^2 {\cal N} (1 + 2 \varphi)} \left(
       \chi^{i} \kappa_{,i}
       + c {\varphi^{,i} {\cal N}_{,i} \over 1 + 2 \varphi} \right)
       + c^2 \overline{K}^i_j \overline{K}^j_i.
   \label{eq4}
\eea
Tracefree ADM propagation:
\bea
   & &
       \left( {1 \over {\cal N}} {\partial \over \partial t}
       + 3 H - \kappa + {c \chi^{k} \over a^2 {\cal N} (1 + 2 \varphi)} \nabla_k \right)
   \nonumber \\
   & & \qquad
       \times
       \left\{ {c \over a^2 {\cal N} (1 + 2 \varphi)} \left[
       {1 \over 2} \left( \chi^i_{\;\;,j} + \chi_j^{\;,i} \right)
       - {1 \over 3} \delta^i_j \chi^k_{\;\;,k}
       - {1 \over 1 + 2 \varphi} \left( \chi^{i} \varphi_{,j}
       + \chi_{j} \varphi^{,i}
       - {2 \over 3} \delta^i_j \chi^{k} \varphi_{,k} \right)
       \right] \right\}
   \nonumber \\
   & & \qquad
       - {c^2 \over a^2 ( 1 + 2 \varphi)}
       \left[ {1 \over 1 + 2 \varphi}
       \left( \nabla^i \nabla_j - {1 \over 3} \delta^i_j \Delta \right) \varphi
       + {1 \over {\cal N}}
       \left( \nabla^i \nabla_j - {1 \over 3} \delta^i_j \Delta \right) {\cal N} \right]
   \nonumber \\
   & & \qquad
       =
       {8 \pi G \over c^4} {\widetilde \mu + \widetilde p \over 1 + 2 \varphi} \left(
       v^{i}v_{j}
       - {1 \over 3} \delta^i_j v^{k} v_{k} \right)
       + {c^2 \over a^4 {\cal N}^2 (1 + 2 \varphi)^2} \Bigg[
       {1 \over 2} \left( \chi^{i,k} \chi_{j,k}
       - \chi_{k,j} \chi^{k,i} \right)
   \nonumber \\
   & & \qquad
       + {1 \over 1 + 2 \varphi} \left(
       \chi^{k,i} \chi_k \varphi_{,j}
       - \chi^{i,k} \chi_j \varphi_{,k}
       + \chi_{k,j} \chi^k \varphi^{,i}
       - \chi_{j,k} \chi^i \varphi^{,k} \right)
       + {2 \over (1 + 2 \varphi)^2} \left(
       \chi^{i} \chi_{j} \varphi^{,k} \varphi_{,k}
       - \chi^{k} \chi_{k} \varphi^{,i} \varphi_{,j} \right) \Bigg]
   \nonumber \\
   & & \qquad
       - {c^2 \over a^2 (1 + 2 \varphi)^2}
       \left[ {3 \over 1 + 2 \varphi}
       \left( \varphi^{,i} \varphi_{,j}
       - {1 \over 3} \delta^i_j \varphi^{,k} \varphi_{,k} \right)
       + {1 \over {\cal N}} \left(
       \varphi^{,i} {\cal N}_{,j}
       + \varphi_{,j} {\cal N}^{,i}
       - {2 \over 3} \delta^i_j \varphi^{,k} {\cal N}_{,k} \right) \right].
   \label{eq5}
\eea
ADM energy conservation:
\bea
   & & {1 \over {\cal N}} \left( {\partial \over \partial t}
       + {c \chi^{i} \over a^2 (1 + 2 \varphi)} \nabla_i \right)
       \left[ \widetilde \mu
       + \left( \widetilde \mu + \widetilde p \right)
       \left( \widehat \gamma^2 - 1 \right) \right]
       + \left( \widetilde \mu + \widetilde p \right)
       \left( 3 H - \kappa \right) {1 \over 3} \left(
       4 \widehat \gamma^2 - 1 \right)
   \nonumber \\
   & & \qquad
       + \left( \nabla_i
       + {3 \varphi_{,i} \over 1 + 2 \varphi}
       + 2 {{\cal N}_{,i} \over {\cal N}} \right)
       \left( {\widetilde \mu + \widetilde p \over a (1 + 2 \varphi)}
       \widehat \gamma v^{i} \right)
   \nonumber \\
   & & \qquad
       = - {\widetilde \mu + \widetilde p \over c a^2 {\cal N} (1 + 2 \varphi)^2}
       \left[ \chi^{i,j} v_{i} v_{j}
       - {1 \over 3} \chi^j_{\;\;,j} v^{i} v_{i}
       - {2 \over 1 + 2 \varphi} \left(
       v^{i} v^{j} \chi_{i} \varphi_{,j}
       - {1 \over 3} v^{i} v_{i} \chi^{j} \varphi_{,j} \right) \right].
   \label{eq6}
\eea
ADM momentum conservation:
\bea
   & & \left( {1 \over {\cal N}} {\partial \over \partial t}
       + 3 H - \kappa + {c \chi^{j} \over a^2 {\cal N} (1 + 2 \varphi)} \nabla_j \right)
       \left[
       a \left( \widetilde \mu + \widetilde p \right) \widehat \gamma v_{i} \right]
       + c^2 \widetilde p_{,i}
       + c^2 \left( \widetilde \mu + \widetilde p \right) {{\cal N}_{,i} \over {\cal N}}
       =
       - \left[ \left( \widetilde \mu + \widetilde p \right)
       {v^{j} v_{i} \over 1 + 2 \varphi}
       \right]_{,j}
   \nonumber \\
   & & \qquad
       - {c \over a {\cal N}}
       \left( {\chi^{j} \over 1 + 2 \varphi} \right)_{,i}
       \left( \widetilde \mu + \widetilde p \right) \widehat \gamma v_{j}
       - {\widetilde \mu + \widetilde p \over 1 + 2 \varphi}
       v^{j} \left[ {1 \over 1 + 2 \varphi}
       \left( 3 v_{i} \varphi_{,j}
       - v_{j} \varphi_{,i} \right)
       + {1 \over {\cal N}}
       \left( v_{i} {\cal N}_{,j}
       + v_{j} {\cal N}_{,i} \right) \right].
   \label{eq7}
\eea Equation (\ref{eq1}) follows from the definition of $\kappa$ as
$K \equiv - 3 H + \kappa$; $K$ is the trace of extrinsic curvature
presented in equation (\ref{extrinsic-curvature}). Equations
(\ref{eq2})-(\ref{eq7}) follow from the ADM equations in equations
(\ref{E-constraint})-(\ref{Mom-conservation}). We
used the Lorentz factor \bea
   & & \widehat \gamma \equiv
       \sqrt{ 1 + {v^k v_k \over c^2 (1 + 2 \varphi)} },
\eea introduced in equation (\ref{Lorentz-factor}). In the Appendix
\ref{sec:velocities} we have introduced more physically
motivated fluid three velocities, $\widehat v^i$ and $\overline
v^i$; $\widehat v^i$ is the fluid three-velocity measured by the
Eulerian observer, and $\overline v^i$ is the fluid coordinate
three-velocity. The relations among the three definitions of fluid
three-velocity are presented in equation (\ref{v-relation}). The
variable ${\cal N}$ is related to the lapse function in equation
(\ref{ADM-metric}), and $\overline{K}^i_j$ is the tracefree part of
extrinsic curvature in equation (\ref{extrinsic-curvature}). With
${\cal N}$ and $\overline{K}^i_j \overline{K}^j_i$ given as \bea
   & & {\cal N} = \sqrt{ 1 + 2 \alpha + {\chi^{k} \chi_{k} \over
       a^2 (1 + 2 \varphi)}}, \quad
       \overline{K}^i_j \overline{K}^j_i
       = {1 \over a^4 {\cal N}^2 (1 + 2 \varphi)^2}
       \Bigg\{
       {1 \over 2} \chi^{i,j} \left( \chi_{i,j} + \chi_{j,i} \right)
       - {1 \over 3} \chi^i_{\;\;,i} \chi^j_{\;\;,j}
   \nonumber \\
   & &
       - {4 \over 1 + 2 \varphi} \left[
       {1 \over 2} \chi^i \varphi^{,j} \left(
       \chi_{i,j} + \chi_{j,i} \right)
       - {1 \over 3} \chi^i_{\;\;,i} \chi^j \varphi_{,j} \right]
       + {2 \over (1 + 2 \varphi)^2} \left(
       \chi^{i} \chi_{i} \varphi^{,j} \varphi_{,j}
       + {1 \over 3} \chi^i \chi^j \varphi_{,i} \varphi_{,j} \right) \Bigg\},
   \label{K-bar-eq}
\eea equations (\ref{eq1})-(\ref{eq7}) are the complete set of fully
nonlinear perturbation equations valid for the scalar- and
vector-type perturbations assuming ideal fluid in a flat background.

Instead of equations (\ref{eq6}) and (\ref{eq7}) based on the ADM
equations we can use alternative forms based on the covariant
equations. Equations (\ref{covariant-E-conserv}) and
(\ref{covariant-mom-conserv}) give the following covariant conservation equations.

\noindent
Covariant energy conservation:
\bea
   & & \left[
       {\partial \over \partial t}
       + {1 \over a (1 + 2 \varphi)} \left( {{\cal N} \over \widehat \gamma} v^i
       + {c \over a} \chi^i \right)
       \nabla_i \right] \widetilde \mu
   \nonumber \\
   & & \qquad
       + \left( \widetilde \mu + \widetilde p \right)
       \left[ \left( 3 H - \kappa \right) {\cal N}
       + {({\cal N} v^i)_{,i} \over a \widehat \gamma (1 + 2 \varphi)}
       + {{\cal N} v^i \varphi_{,i} \over
       a \widehat \gamma (1 + 2 \varphi)^2}
       + {1 \over \widehat \gamma} \left(
       {\partial \over \partial t}
       + {c \chi^i \over a^2 (1 + 2 \varphi)} \nabla_i \right)
       \widehat \gamma \right]
       = 0.
   \label{eq6-cov}
\eea
Covariant momentum conservation:
\bea
   & & \left[
       {\partial \over \partial t}
       + {1 \over a (1 + 2 \varphi)}
       \left( {{\cal N} \over \widehat \gamma} v^k
       + {c \over a} \chi^k \right)
       \nabla_k \right]
       a v_i
       + {1 \over \widetilde \mu + \widetilde p} \left\{
       c^2 {{\cal N} \over \widehat \gamma} \widetilde p_{,i}
       + a v_i \left[
       {\partial \over \partial t}
       + {1 \over a (1 + 2 \varphi)}
       \left( {{\cal N} \over \widehat \gamma} v^k
       + {c \over a} \chi^k \right)
       \nabla_k \right]  \widetilde p
       \right\}
   \nonumber \\
   & & \qquad
       + c^2 \widehat \gamma {\cal N}_{,i}
       + {1 - \widehat \gamma^2 \over \widehat \gamma}
       {c^2 {\cal N} \varphi_{,i} \over 1 + 2 \varphi}
       + {c \over a}
       v^k \nabla_i \left( {\chi_k \over 1 + 2 \varphi} \right)
       = 0.
   \label{eq7-cov}
\eea These are alternative
forms of equations (\ref{eq6}) and (\ref{eq7}), respectively. For
comparison between the ADM and covariant conservation equations, see
equation (\ref{conservation-cov-ADM}). We can consider this set of
equations as exact, or treat it perturbatively to the fully
nonlinear order. The perturbation variables are $\delta \mu$, $v_i$,
$\kappa$, $\chi_i$, $\varphi$ and $\alpha$; $\delta p$ should be
provided by an equation of state. Notice that we have not separated
the background order equations yet; we only have assumed that $a$ is
a function of time. The vector-type perturbation is contained in
$v_i$ and $\chi_i$ as \bea
   & & v_i \equiv - v_{,i} + v^{(v)}_i, \quad
       \chi_i \equiv c \chi_{,i} + a \Psi^{(v)}_i.
\eea For the vector-type perturbation, equations (\ref{eq3}) and (\ref{eq7}) to the linear order give equations (\ref{eq3-vector}) and (\ref{eq7-vector}), respectively. For the pure scalar-type perturbation we set $v^{(v)}_i = 0 = \Psi^{(v)}_i$, thus $v_i = - v_{,i}$ and $\chi_i = c \chi_{,i}$. The dimensions are \bea
   & & [a] = [\widetilde g_{ab}] = [\widetilde u_a]
       = [\alpha] = [\varphi] = [\chi^i] = [\Psi^{(v)}_i]
       = [v^i/c]
       = [\widehat v^i/c]
       = [\overline v^i/c]
       = [\widehat \gamma] = 1, \quad
       [v/c] = L,
   \nonumber \\
   & &
       [x^i] = [ c dt] \equiv [ a d \eta] = L, \quad
       [\chi] = T, \quad
       [\kappa] = T^{-1}, \quad
       [\widetilde T_{ab}] = [\widetilde \mu]
       = [\widetilde \varrho c^2]
       = [\widetilde p], \quad
       [G \widetilde \varrho] = T^{-2}, \quad
       [\Lambda] = L^{-2}.
\eea

In the above set of equations we have not taken the temporal gauge condition yet. In a sense the equations are in a sort of gauge-ready form. As the temporal gauge condition we can impose any one condition in equation (\ref{temporal-gauges}), except for the synchronous gauge which leaves the remnant gauge mode; see explanation in the next paragraph below equation (\ref{GI}). Thus, as the gauge conditions we have
\bea
   & & {\rm comoving \; gauge:}              \hskip 2.0cm     v \equiv 0,
   \nonumber \\
   & & {\rm zero\!-\!shear \; gauge:}        \hskip 1.72cm     \chi \equiv 0,
   \nonumber \\
   & & {\rm uniform\!-\!curvature \; gauge:} \hskip .56cm      \varphi \equiv 0,
   \nonumber \\
   & & {\rm uniform\!-\!expansion \; gauge:} \hskip .52cm      \kappa \equiv 0,
   \nonumber \\
   & & {\rm uniform\!-\!density \; gauge:}   \hskip .93cm      \delta \equiv 0,
   \label{temporal-gauges-NL}
\eea now valid to all perturbation orders. These are the fundamental
gauge conditions available to the fully nonlinear order. Also
available ones as the gauge conditions are setting any
non-gauge-invariant linear combination of these fundamental gauge
conditions equal to zero. We can also take the different gauge
condition for the different perturbation order. In these ways we
have infinite number of gauge conditions available, which was true
even to the linear order, now to each perturbation order. Under
these gauge conditions which remove the gauge mode completely, all
perturbation variables have the unique gauge-invariant counterparts,
thus we can identify these as the gauge-invariant variables.
Therefore, the nonlinear perturbation variables in any of our
suggested gauge conditions mentioned above can be regarded as
gauge-invariant ones.

In the Appendix \ref{sec:Multiple} we will present the case of multiple-component
fluid system. The above set of equations remains valid with the
fluid quantities interpreted as the collective ones. We will present
the relations of the collective fluid quantities with the individual
one, see equations (\ref{fluid-Multi-1}) and (\ref{fluid-Multi-2}),
and present the additional energy and the momentum conservation
equations of the individual component, see equations
(\ref{I-1})-(\ref{I-4}) and the prescription explained above
equation (\ref{I-1}).

In the Appendix \ref{sec:MSF} we will present the case of a minimally coupled scalar field. The fluid equations in this section are valid with the fluid quantities expressed in terms of the scalar field in equation (\ref{fluids-MSF}). We additionally have the equation of motion of the field in equation (\ref{EOM}).

\section{Third-order perturbation equations in a gauge-ready form}
                                             \label{sec:3rd-order}

In the nonlinear perturbation approach we assume the perturbation variables $\delta$, $v_i$, $\kappa$, $\chi_i$, $\alpha$ and $\varphi$ are small. As one example, here we present pure scalar-type perturbation equations valid to the third order in perturbations without fixing the temporal gauge condition, thus $v_i = - v_{,i}$ and $\chi_i = \chi_{,i}$. Equations up to third-order perturbations are needed to get the leading nonlinear contribution to the power spectrum.

\noindent
Definition of $\kappa$:
\bea
   & & \kappa - 3 H \alpha + 3 \dot \varphi + {\Delta \over a^2} \chi
       = {1 \over 2 a^2} \chi^{,k} \chi_{,k}
       \left( 3 H + 3 \dot \varphi
       - 6 H \varphi
       - 9 H \alpha
       + {\Delta \over a^2} \chi \right)
       - {1 \over a^2} \chi^{,k} \varphi_{,k} \left( 1 - \alpha - 4 \varphi \right)
   \nonumber \\
   & & \qquad
       + {3 \over 2} H \alpha^2 \left( -3 + 5 \alpha \right)
       + \left( 3 \dot \varphi
       + {\Delta \over a^2} \chi \right)
       \left( \alpha + 2 \varphi
       - {3 \over 2} \alpha^2
       - 2 \alpha \varphi
       - 4 \varphi^2 \right).
   \label{eq1-3rd}
\eea
ADM energy constraint:
\bea
   & & - {3 \over 2} \left( H^2 - {8 \pi G \over 3} \mu - {\Lambda \over 3} \right)
       + 4 \pi G \delta \mu + H \kappa + {\Delta \over a^2} \varphi
       = {1 \over 6} \kappa^2
       - 4 \pi G \left( \mu + p \right) v^{,i} v_{,i}
       \left( 1 - 2 \varphi + {\delta \mu + \delta p \over \mu + p} \right)
       + 4 \left( {\Delta \over a^2} \varphi \right) \varphi
       \left( 1 - 3 \varphi \right)
   \nonumber \\
   & & \qquad
       + {3 \over 2 a^2} \varphi^{,i} \varphi_{,i} \left( 1 - 6 \varphi \right)
       - {1 \over 4 a^4} \left\{
       \left[ \chi^{,ij} \chi_{,ij}
       - {1 \over 3} \left( \Delta \chi \right)^2 \right]
       \left( 1 - 2 \alpha - 4 \varphi \right)
       - 4 \chi^{,ij} \chi_{,i} \varphi_{,j}
       + {4 \over 3} \left( \Delta \chi \right) \chi^{,i} \varphi_{,i}
       \right\}.
   \label{eq2-3rd}
\eea
ADM momentum constraint:
\bea
   & & {2 \over 3} \left[ \kappa
       + {\Delta \over a^2} \chi
       - 12 \pi G \left( \mu + p \right) a v \right]_{,i}
       = 8 \pi G \left( \mu + p \right) a v_{,i}
       \left( {\delta \mu + \delta p \over \mu + p}
       + {1 \over 2} v^{,k} v_{,k} \right)
   \nonumber \\
   & & \qquad
       + {2 \over 3} \left( {\Delta \over a^2} \chi \right)_{,i}
       \left( \alpha +2 \varphi
       - {3 \over 2} \alpha^2
       - 2 \alpha \varphi
       - 4 \varphi^2
       + {1 \over 2 a^2} \chi^{,k} \chi_{,k} \right)
   \nonumber \\
   & & \qquad
       + {1 \over a^2} \left[ \alpha_{,j} \left( 1 - 3 \alpha - 2 \varphi \right)
       - \varphi_{,j} \left( 1 - \alpha - 4 \varphi \right)
       + {1 \over a^2} \chi^{,k} \chi_{,kj} \right]
       \left( \nabla^j \nabla_i
       - {1 \over 3} \delta^j_i \Delta \right) \chi
   \nonumber \\
   & & \qquad
       + {1 \over a^2}
       \left[ \left( \Delta \chi \right) \varphi_{,i}
       + \chi_{,i} \Delta \varphi
       + {1 \over 3} \left( \chi^{,k} \varphi_{,k} \right)_{,i} \right]
       \left( 1 - \alpha - 4 \varphi \right)
   \nonumber \\
   & & \qquad
       - {1 \over a^2}
       \left[ \chi_{,i} \varphi^{,k} \left( \alpha_{,k} + \varphi_{,k} \right)
       + {1 \over 3} \chi^{,k} \left(
       \varphi_{,k} \varphi_{,i}
       + 3 \varphi_{,i} \alpha_{,k}
       - 2 \varphi_{,k} \alpha_{,i} \right) \right].
   \label{eq3-3rd}
\eea
Trace of ADM propagation:
\bea
   & & - 3 \dot H - 3 H^2
       - 4 \pi G \left( \mu + 3 p \right) + \Lambda
       + \dot \kappa + 2 H \kappa
       - 4 \pi G \left( \delta \mu + 3 \delta p \right)
       + \left( 3 \dot H + {\Delta \over a^2} \right) \alpha
   \nonumber \\
   & & \qquad
       = \dot \kappa \left( \alpha - {3 \over 2} \alpha^2
       + {1 \over 2 a^2} \chi^{,k} \chi_{,k} \right)
       + {1 \over 3} \kappa^2
       + 8 \pi G \left( \mu + p \right) v^{,i} v_{,i}
       \left( 1 - 2 \varphi
       + {\delta \mu + \delta p \over \mu + p} \right)
   \nonumber \\
   & & \qquad
       + {3 \over 2} \dot H \left[
       3 \alpha^2
       - {1 \over a^2} \chi^{,k} \chi_{,k}
       \left( 1 - 3 \alpha - 2 \varphi \right)
       - 5 \alpha^3 \right]
       + \left( \alpha + 2 \varphi
       - {3 \over 2} \alpha^2
       - 2 \alpha \varphi - 4 \varphi^2
       + {1 \over 2 a^2} \chi^{,k} \chi_{,k} \right)
       {\Delta \over a^2} \alpha
   \nonumber \\
   & & \qquad
       + \left( 1 - \alpha - 2 \varphi \right)
       {\Delta \over 2 a^2} \left[
       \alpha^2
       - {1 \over a^2} \chi^{,k} \chi_{,k} \left( 1 - \alpha - 2 \varphi \right)
       - \alpha^3 \right]
       - {1 \over a^2} \left[
       \chi^{,i} \kappa_{,i} \left( 1 - \alpha - 2 \varphi \right)
       + \varphi^{,i} \alpha_{,i} \left( 1 - 2 \alpha - 4 \varphi \right) \right]
   \nonumber \\
   & & \qquad
       + {1 \over a^4} \left\{
       \left[ \chi^{,ij} \chi_{,ij}
       - {1 \over 3} \left( \Delta \chi \right)^2 \right]
       \left( 1 - 2 \alpha - 4 \varphi \right)
       - 5 \chi^{,ij} \chi_{,i} \varphi_{,j}
       + {4 \over 3} \left( \Delta \chi \right) \chi^{,i} \varphi_{,i}
       \right\}.
   \label{eq4-3rd}
\eea
Tracefree ADM propagation:
\bea
   & & {1 \over a^2} \left( \nabla^i \nabla_j
       - {1 \over 3} \delta^i_j \Delta \right)
       \left( \dot \chi + H \chi - \alpha - \varphi \right)
       = \left( {\partial \over \partial t}
       + 3 H \right) \Bigg\{ {1 \over a^2} \Bigg[
       \left( \alpha + 2 \varphi
       - {3 \over 2}\alpha^2 - 2 \alpha \varphi
       - 4 \varphi^2
       + {1 \over 2 a^2} \chi^{,k} \chi_{,k} \right)
   \nonumber \\
   & & \qquad
       \times
       \left( \nabla^i \nabla_j - {1 \over 3} \delta^i_j \Delta \right) \chi
       +
       \left( \chi^{,i} \varphi_{,j}
       + \chi_{,j} \varphi^{,i}
       - {2 \over 3} \delta^i_j \chi^{,k} \varphi_{,k} \right)
       \left( 1 - \alpha - 4 \varphi \right)
       \Bigg] \Bigg\}
   \nonumber \\
   & & \qquad
       + \left[ \left( \alpha - {3 \over 2} \alpha^2
       + {1 \over 2 a^2} \chi^{,k} \chi_{,k} \right) {\partial \over \partial t}
       + \kappa
       - {1 \over a^2} \left( 1 - \alpha - 2 \varphi \right)
       \chi^{,k} \nabla_k \right]
       \Bigg\{ {1 \over a^2} \Bigg[
       \left( 1 - \alpha - 2 \varphi \right) \left( \nabla^i \nabla_j
       - {1 \over 3} \delta^i_j \Delta \right) \chi
   \nonumber \\
   & & \qquad
       - \left( \chi^{,i} \varphi_{,j}
       + \chi_{,j} \varphi^{,i}
       - {2 \over 3} \delta^i_j \chi^{,\ell} \varphi_{,\ell} \right)
       \Bigg] \Bigg\}
       - {1 \over a^4} \varphi^{,k} \left( \chi^{,i}_{\;\;\; k} \chi_{,j}
       + \chi_{,jk} \chi^{,i}
       - {2 \over 3} \delta^i_j \chi_{,k\ell} \chi^{,\ell} \right)
   \nonumber \\
   & & \qquad
       - {1 \over a^2} \left( \alpha + 2 \varphi
       - {3 \over 2} \alpha^2
       - 2 \alpha \varphi
       - 4 \varphi^2
       + {1 \over 2 a^2} \chi^{,k} \chi_{,k} \right)
       \left( \nabla^i \nabla_j - {1 \over 3} \delta^i_j \Delta \right)
       \alpha
   \nonumber \\
   & & \qquad
       + {1 \over 2 a^2} \left( 1 - \alpha - 2 \varphi \right)
       \left( \nabla^i \nabla_j - {1 \over 3} \delta^i_j \Delta \right)
       \left[ - \alpha^2
       + {1 \over a^2} \chi^{,k} \chi_{,k}
       \left( 1 - \alpha - 2 \varphi \right) + \alpha^3 \right]
   \nonumber \\
   & & \qquad
       - {3 \over a^2}
       \left( \varphi^{,i} \varphi_{,j}
       - {1 \over 3} \delta^i_j \varphi^{,k} \varphi_{,k} \right)
       \left( 1 - 6 \varphi \right)
       - {1 \over a^2}
       \left( \varphi^{,i} \alpha_{,j}
       + \varphi_{,j} \alpha^{,i}
       - {2 \over 3} \delta^i_j \varphi^{,k} \alpha_{,k} \right)
       \left( 1 - 2 \alpha - 4 \varphi \right)
   \nonumber \\
   & & \qquad
       - {4 \over a^2} \varphi \left(
       1 - 3 \varphi \right)
       \left( \nabla^i \nabla_j - {1 \over 3} \delta^i_j \Delta \right) \varphi
       + 8 \pi G \left( \mu + p \right)
       \left( v^{,i} v_{,j}
       - {1 \over 3} \delta^i_j v^{,k} v_{,k} \right)
       \left( 1 - 2 \varphi
       + {\delta \mu + \delta p \over \mu + p} \right).
   \label{eq5-3rd}
\eea
ADM energy conservation:
\bea
   & & \dot \mu + 3 H \left( \mu + p \right)
       + \delta \dot \mu + 3 H \left( \delta \mu + \delta p \right)
       - \left( \mu + p \right) \kappa - \dot \mu \alpha
       - \left( \mu + p \right) {\Delta \over a} v
       = \left( \delta \mu + \delta p \right) \kappa
       + {4 \over 3} \left( \mu + p \right) \kappa v^{,k} v_{,k}
   \nonumber \\
   & & \qquad
       + {1 \over 2} \dot \mu
       \left[ - 3 \alpha^2
       + {1 \over a^2} \chi^{,k} \chi_{,k}
       \left( 1 - 3 \alpha - 2 \varphi \right)
       + 5 \alpha^3 \right]
       + \delta \dot \mu \left( \alpha
       - {3 \over 2} \alpha^2 + {1 \over 2 a^2} \chi^{k} \chi_{,k} \right)
       - {1 \over a^2} \chi^{,i} \delta \mu_{,i}
       \left( 1 - \alpha - 2 \varphi \right)
   \nonumber \\
   & & \qquad
       - \left( 1 - \alpha \right) {\partial \over \partial t} \left[
       \left( \mu + p \right) v^{,k} v_{,k}
       \left( 1 - 2 \varphi + {\delta \mu + \delta p \over \mu + p} \right)
       \right]
       - 4 H \left( \mu + p \right) v^{,k} v_{,k}
       \left( 1 - 2 \varphi + {\delta \mu + \delta p \over \mu + p} \right)
   \nonumber \\
   & & \qquad
       + {1 \over a} \left( \mu + p \right)
       \nabla_i \left\{ v^{,i} \left[
       - 2 \varphi + {\delta \mu + \delta p \over \mu + p} \left( 1 - 2 \varphi \right)
       + 4 \varphi^2 + {1 \over 2} v^{,k} v_{,k} \right] \right\}
   \nonumber \\
   & & \qquad
       + {1 \over a} \left( \mu + p \right) v^{,i}
       \left( 1 + {\delta \mu + \delta p \over \mu + p} \right)
       \left[ 3 \varphi_{,i} \left( 1 - 4 \varphi \right)
       + 2 \alpha_{,i} \left( 1 - 2 \alpha - 2 \varphi \right) \right]
   \nonumber \\
   & & \qquad
       + {1 \over a^2} \left( \mu + p \right)
       v^{,i} \left[
       {2 \over a} \chi_{,ij} \chi^{,j}
       - 2 v_{,ij} \chi^{,j}- \chi_{,ij} v^{,j}
       + {1 \over 3} \left( \Delta \chi \right) v_{,i} \right].
   \label{eq6-3rd}
\eea
ADM momentum conservation:
\bea
   & & \left\{ { 1\over a^3} \left[ a^4 \left( \mu + p \right) v
       \right]^{\displaystyle\cdot}
       - \left( \mu + p \right) \alpha
       - \delta p \right\}_{,i}
       = - \left( {\partial \over \partial t} + 3 H \right)
       \left[ a \left( \mu + p \right) v_{,i}
       \left( {\delta \mu + \delta p \over \mu + p}
       + {1 \over 2} v^{,k} v_{,k} \right) \right]
   \nonumber \\
   & & \qquad
       + \left[ \left( \alpha - {3 \over 2} \alpha^2
       + {1 \over 2 a^2} \chi^{,k} \chi_{,k} \right)
       {\partial \over \partial t}
       + \kappa
       - {1 \over a^2} \left( 1 - \alpha - 2 \varphi \right)
       \chi^{,k} \nabla_k \right] \left[ a \left(\mu + p \right) v_{,i}
       \left( 1 + {\delta \mu + \delta p \over \mu + p} \right) \right]
   \nonumber \\
   & & \qquad
   + \left[ \left(\mu + p \right) v^{,k} v_{,i}
       \left( 1 - 2 \varphi + {\delta \mu + \delta p \over \mu + p} \right)
       \right]_{,k}
       - {1 \over a} \left(\mu + p \right) v_{,k}
       \left( 1 + {\delta \mu + \delta p \over \mu + p} \right)
       \left[ \chi^{,k}_{\;\;\; i} \left( 1 - \alpha - 2 \varphi \right)
       - 2 \chi^{,k} \varphi_{,i} \right]
   \nonumber \\
   & & \qquad
       + \left( \mu + p \right) v^{,k}
       \left[ v_{,i} \left( \alpha + 3 \varphi \right)_{,k}
       + v_{,k} \left( \alpha - \varphi \right)_{,i} \right]
   \nonumber \\
   & & \qquad
       + \left( \mu + p \right) \left\{
       \alpha_{,i} \left[ - 2 \alpha + {\delta \mu + \delta p \over \mu + p}
       \left( 1 - 2 \alpha \right) + 4 \alpha^2 \right]
       + {1 \over 2 a^2} \left( 1 + {\delta \mu + \delta p \over \mu + p} \right)
       \left[ \chi^{,k} \chi_{,k}
       \left( 1 - 2 \alpha - 2 \varphi \right) \right]_{,i} \right\}.
   \label{eq7-3rd}
\eea

To the background order, equations (\ref{eq2-3rd}), (\ref{eq4-3rd}) and (\ref{eq6-3rd}), respectively, give \bea
   & & H^2 = {8 \pi G \over 3} \mu + {\Lambda \over 3}, \quad
       {\ddot a \over a}
       = - {4 \pi G \over 3} \left( \mu + 3 p \right)
       + {\Lambda \over 3}, \quad
       \dot \mu + 3 H \left( \mu + p \right) = 0.
\eea To the linear order, assuming the background equations separately, equations (\ref{eq1-3rd})-(\ref{eq7-3rd}) give equations (\ref{eq1-linear})-(\ref{eq7-linear}).

The equations of motion of the scalar field to the background order and to the third-order perturbation are presented in equations (\ref{EOM-BG-order}) and (\ref{EOM-third-order}), respectively.

\section{Comoving gauge}
                                             \label{sec:CG}

In sections \ref{sec:CG} and \ref{sec:other-gauges} we consider only the scalar-type perturbation, thus $v_i = - v_{,i}$ and $\chi_i = \chi_{,i}$.
In the comoving gauge we set \bea
   & & v \equiv 0.
\eea
We have $v_i = 0$, thus the fluid four-vector becomes a normal one with $\widetilde u_i = 0$. Equation (\ref{eq7}) gives \bea
   & & \widetilde p_{,i}
       = - \left( \widetilde \mu + \widetilde p \right) {{\cal N}_{,i} \over {\cal N}}.
   \label{eq7-CG}
\eea
Equations (\ref{eq1})-(\ref{eq7}) give a (redundantly) complete set of equations for the variables $\delta$, $\kappa$, $\varphi$, $\chi$ and $\alpha$. We can treat this set of equation either exactly or perturbatively to all orders. As explained in Section \ref{sec:convention} and below equation (\ref{temporal-gauges-NL}), all perturbation variables in the comoving gauge are gauge invariant to the nonlinear order as \bea
   & & \delta = \delta_v, \quad
      \kappa = \kappa_v, \quad
      \varphi = \varphi_v, \quad
      \chi = \chi_v, \quad
      \alpha = \alpha_v.
\eea

We have several ways of having closed form second order differential equations.
Equations (\ref{eq6}) and (\ref{eq4}), together with equations (\ref{eq2}), (\ref{eq3}) and (\ref{eq7}) to determine $\alpha$, $\varphi$ and $\chi$, give equations for $\dot \delta$ and $\dot \kappa$.
Equations (\ref{eq1}) and (\ref{eq4}), together with equations (\ref{eq2}), (\ref{eq3}) and (\ref{eq7}) to determine $\delta$, $\alpha$ and $\chi$, give equations for $\dot \varphi$ and $\dot \kappa$. Etc.
For example, assuming the background equations are valid separately, equation (\ref{eq6}) gives \bea
   & & \kappa = {1 \over \widetilde \mu + \widetilde p} {1 \over {\cal N}}
       \left( {\partial \over \partial t}
       + {\chi^{,i} \over a^2 ( 1 + 2 \varphi )} \nabla_i \right)
       {\widetilde \mu}
       - {\dot \mu \over \mu + p}.
\eea By removing $\kappa$ in equation (\ref{eq4}) we have a second-order differential equation for $\delta$ as \bea
   & & {1 \over {\cal N}}
       \left[ {1 \over \widetilde \mu + \widetilde p}
       {1 \over {\cal N}}
       \left( {\partial \over \partial t}
       + {\chi^{,i} \over a^2 ( 1 + 2 \varphi )} \nabla_i \right)
       {\widetilde \mu}
       - {\dot \mu \over \mu + p} \right]^{\displaystyle\cdot}
       + 2 H
       \left[ {1 \over \widetilde \mu + \widetilde p}
       {1 \over {\cal N}}
       \left( {\partial \over \partial t}
       + {\chi^{,i} \over a^2 ( 1 + 2 \varphi )} \nabla_i \right)
       {\widetilde \mu}
       - {\dot \mu \over \mu + p} \right]
   \nonumber \\
   & & \qquad
       - 3 \left( {1 \over {\cal N}} - 1 \right) \dot H
       - 4 \pi G \left( \delta \mu + 3 \delta p \right)
       + {\Delta {\cal N} \over a^2 {\cal N} (1 + 2 \varphi)}
       = {1 \over 3} \kappa^2
       - {1 \over a^2 {\cal N} (1 + 2 \varphi)} \left(
       \chi^{i} \kappa_{,i}
       + {\varphi^{,i} {\cal N}_{,i} \over 1 + 2 \varphi} \right)
       + \overline{K}^i_j \overline{K}^j_i.
\eea The similar equation can be derived from equation
(\ref{Jackson-eq}) evaluated for $v_i = 0$.

If we have a solution for a given variable we can derive the rest of the variables using the above complete set of equations. From these solutions we can derive all the variables in any other gauge conditions using the gauge-ready form equations in (\ref{eq1})-(\ref{eq7}) and explicit construction of gauge-invariant combinations; for solutions in the matter dominated era to the second order perturbations, see Hwang, et al (2012).

\subsection{Zero-pressure case}

For $\widetilde p = 0$, equation (\ref{eq7-CG}) gives ${\cal N}_{,i} = 0$. Thus, we can set ${\cal N} = 1$ with \bea
   & & \alpha
       = - {1 \over 2} {\chi^{,k} \chi_{,k} \over a^2 (1 + 2 \varphi)}.
   \label{alpha-CG} \\
   \nonumber
\eea Thus, in our spatial gauge condition with $\gamma = 0$, the comoving gauge ($v = 0$) is no longer synchronous ($\alpha = 0$) to the nonlinear order  even in the zero-pressure medium (Hwang \& Noh 2006).

Equations (\ref{eq4}) and (\ref{eq6}) together with equations (\ref{eq2}) and (\ref{eq3}) provide a complete set of equations for $\delta$ and $\kappa$. We have
\bea
   & & \left( {\dot \mu \over \mu} + 3 H \right) \left( 1 + \delta \right)
       +
       \dot \delta - \kappa = \delta \kappa
       - {\chi^{,i} \delta_{,i} \over a^2 (1 + 2 \varphi)},
   \label{eq6-CG-mde} \\
   & & - 3 \dot H - 3 H^2 - 4 \pi G \mu + \Lambda
       +
       \dot \kappa
       + 2 H \kappa
       - 4 \pi G \delta \mu
       = {1 \over 3} \kappa^2
       - {\chi^{,i} \kappa_{,i} \over a^2 (1 + 2 \varphi)}
       + \overline{K}^i_j \overline{K}^j_i,
   \label{eq4-CG-mde}
\eea
with $\chi$ and $\varphi$ determined by
\bea
   & & \kappa_{,i}
       + {\Delta \chi_{,i} \over a^2 ( 1 + 2 \varphi )}
       =
       {1 \over a^2 ( 1 + 2 \varphi)^2}
       \left[
       2 \left( \Delta \chi \right) \varphi_{,i}
       + {1 \over 2} \chi^{,k} \varphi_{,ik}
       - \chi_{,ik} \varphi^{,k}
       + {3 \over 2} \chi_{,i} \Delta \varphi
       - {3 \over 2} {1 \over 1 + 2 \varphi}
       \left( \chi_{,i} \varphi_{,k}
       + {1 \over 3} \chi_{,k} \varphi_{,i} \right) \varphi^{,k} \right],
   \label{eq3-CG-mde} \\
   & & - {3 \over 2} \left( H^2
       - {8 \pi G \over 3} \mu - {\Lambda \over 3} \right)
       +
       H \kappa
       + 4 \pi G \mu \delta
       + {\Delta \varphi \over a^2 (1 + 2 \varphi)^2}
       = {1 \over 6} \kappa^2
       + {3 \over 2} {\varphi^{,i} \varphi_{,i} \over a^2 (1 + 2 \varphi)^3}
       - {1 \over 4} \overline{K}^i_j \overline{K}^j_i,
   \label{eq2-CG-mde}
\eea
where \bea
   & & \overline{K}^i_j \overline{K}^j_i
       = {1 \over a^4 (1 + 2 \varphi)^2}
   \nonumber \\
   & & \qquad
       \times
       \left\{
       \chi^{,ij} \chi_{,ij}
       - {1 \over 3} \left( \Delta \chi \right)^2
       + {4 \over 1 + 2 \varphi} \left[
       {1 \over 3 } \left( \Delta \chi \right) \chi^{,i} \varphi_{,i}
       - \chi^{,ij} \chi_{,i} \varphi_{,j} \right]
       + {2 \over (1 + 2 \varphi)^2} \left[
       {1 \over 3} \left( \chi^{,i} \varphi_{,i} \right)^2
       + \chi^{,i} \chi_{,i} \varphi^{,j} \varphi_{,j} \right] \right\}.
   \label{KK-CG-mde}
\eea These equations are still exact. {\it Assuming} the background equations are valid separately, from equations (\ref{eq6-CG-mde}) and (\ref{eq4-CG-mde}) together with equations (\ref{eq3-CG-mde}) and (\ref{eq2-CG-mde}), we can derive a closed form second-order differential equation for $\delta$ or $\kappa$. For $\delta$, we have \bea
   & & \ddot \delta
       + 2 H \dot \delta
       - 4 \pi G \mu \delta
       = {1 \over a^2} \left( a^2 \kappa \delta \right)^{\displaystyle\cdot}
       - {1 \over a^2} \left( {\chi^{,i} \delta_{,i} \over 1 + 2 \varphi} \right)^{\displaystyle\cdot}
       + {1 \over 3} \kappa^2
       - {\chi^{,i} \kappa_{,i} \over a^2 (1 + 2 \varphi)}
       + \overline{K}^i_j \overline{K}^j_i.
\eea Using equations (\ref{eq6-CG-mde}), (\ref{eq3-CG-mde}) and (\ref{eq2-CG-mde}) to determine $\kappa$, $\varphi$ and $\chi$, we can (perturbatively) express this equation purely in terms of $\delta$.

Now, the equation for $\dot \varphi$ follows from equations
(\ref{eq1}) and (\ref{eq3-CG-mde}) as \bea
   & & \left[ \ln{ \left( 1 + 2 \varphi \right) } \right]^{\displaystyle\cdot}_{,i}
       = {1 \over a^2 ( 1 + 2 \varphi )^2}
       \left[ \chi^{,k} \varphi_{,ik}
       + \chi_{,i} \Delta \varphi
       - {1 \over 1 + 2 \varphi}
       \left( \chi_{,i} \varphi_{,k}
       + 3 \chi_{,k} \varphi_{,i} \right) \varphi^{,k} \right],
\eea thus $\dot \varphi = 0$ to the linear order. Together with equation (\ref{eq4-CG-mde}), using equations (\ref{eq3-CG-mde}) and (\ref{eq2-CG-mde}) to determine $\delta$ and $\chi$, we have the closed form equations purely in terms of $\varphi$ and $\kappa$.

\subsection{Equations to fifth order in the zero-pressure case}

As one exercise demonstrating the power of our fully nonlinear formulation, we present the perturbation equations to the fifth order. We consider a zero-pressure fluid in the comoving gauge. Obviously the derivation is quite simple requiring only Taylor expansion of $1/(1 + 2 \varphi)$ terms. {\it Assuming} the background equations are valid separately, from equations (\ref{eq6-CG-mde})-(\ref{KK-CG-mde}) we have
\bea
   & &
       \dot \delta - \kappa = \delta \kappa
       - {1 \over a^2} \chi^{,i} \delta_{,i}
       \left( 1 - 2 \varphi + 4 \varphi^2 - 8 \varphi^3 \right),
   \label{eq6-CG-mde-5th} \\
   & &
       \dot \kappa
       + 2 H \kappa
       - 4 \pi G \delta \mu
       = {1 \over 3} \kappa^2
       - {1 \over a^2} \chi^{,i} \kappa_{,i}
       \left( 1 - 2 \varphi + 4 \varphi^2 - 8 \varphi^3 \right)
       + \overline{K}^i_j \overline{K}^j_i,
   \label{eq4-CG-mde-5th} \\
   & &
       {2 \over 3} \left( \kappa
       + {\Delta \over a^2} \chi \right)_{,i}
       =
       {1 \over a^2} \Bigg[
       {4 \over 3} \left( \Delta \chi_{,i} \right)
       \varphi
       \left( 1 - 2 \varphi + 4 \varphi^2 - 8 \varphi^3 \right)
       - \left( \chi^{,j}_{\;\;\;i}
       - {1 \over 3} \delta^j_i \Delta \chi \right) \varphi_{,j}
       \left( 1 - 4 \varphi + 12 \varphi^2 - 32 \varphi^3 \right)
   \nonumber \\
   & & \qquad
       + \left(
       \chi^{,j} \varphi_{,i}
       + \chi_{,i} \varphi^{,j}
       - {2 \over 3} \delta^j_i \chi^{,k} \varphi_{,k} \right)_{,j}
       \left( 1 - 4 \varphi + 12 \varphi^2 - 32 \varphi^3 \right)
       - \left( \chi_{,i} \varphi_{,j}
       + {1 \over 3} \chi_{,j} \varphi_{,i} \right) \varphi^{,j}
       \left( 1 - 6 \varphi + 24 \varphi^2 \right)
       \Bigg],
   \label{eq3-CG-mde-5th} \\
   & &
       H \kappa
       + 4 \pi G \mu \delta
       + {\Delta \over a^2} \varphi
       =
       {1 \over 6} \kappa^2
       + 4 \left( {\Delta \over a^2} \varphi \right) \varphi
       \left( 1 -3 \varphi + 8 \varphi^2 - 20 \varphi^3 \right)
       + {3 \over 2} {1 \over a^2} \varphi^{,i} \varphi_{,i}
       \left( 1 - 6 \varphi + 24 \varphi^2 - 80 \varphi^3 \right)
       - {1 \over 4} \overline{K}^i_j \overline{K}^j_i,
   \label{eq2-CG-mde-5th}
\eea
where \bea
   & & \overline{K}^i_j \overline{K}^j_i
       = {1 \over a^4}
       \Bigg\{
       \left[ \chi^{,ij} \chi_{,ij}
       - {1 \over 3} \left( \Delta \chi \right)^2 \right]
       \left( 1 - 4 \varphi + 12 \varphi^2 - 32 \varphi^3 \right)
       + 4 \left[
       {1 \over 3 } \left( \Delta \chi \right) \chi^{,i} \varphi_{,i}
       - \chi^{,ij} \chi_{,i} \varphi_{,j} \right]
       \left( 1 - 6 \varphi + 24 \varphi^2 \right)
   \nonumber \\
   & & \qquad
       + 2 \left[
       {1 \over 3} \left( \chi^{,i} \varphi_{,i} \right)^2
       + \chi^{,i} \chi_{,i} \varphi^{,j} \varphi_{,j} \right]
       \left( 1 - 8 \varphi \right)
       \Bigg\}.
   \label{KK-CG-mde-5th}
\eea From equations (\ref{eq6-CG-mde-5th}) and
(\ref{eq4-CG-mde-5th}) together with equation (\ref{eq3-CG-mde-5th})
to the fourth order and equation (\ref{eq2-CG-mde-5th}) to the third
order, we can derive a closed form second-order differential
equation for $\delta$ or $\kappa$.

The fifth order perturbation equation is needed to have the
next-to-leading-order nonlinear contribution to the power spectrum.
The leading nonlinear order power spectrum demands the third order
perturbation and the results for the density and velocity power
spectra are presented in Jeong {\it et al} (2011). The fourth order
perturbation equations will be needed to have the
next-to-leading-order nonlinear contribution to the non-Gaussianity.
The leading order non-Gussianity demands the second order
perturbation.

\subsection{Pure Einstein's gravity corrections to fully nonlinear order}

We can arrange equations (\ref{eq6-CG-mde}), (\ref{eq4-CG-mde}) and (\ref{eq3-CG-mde}) in the following forms
\bea
   & &
       \dot \delta - \kappa
       - \delta \kappa
       + {1 \over a^2} \chi^{,i} \delta_{,i}
       = {2 \varphi \chi^{,i} \delta_{,i} \over a^2 (1 + 2 \varphi)},
   \label{eq6-CG-mde-pure-GR} \\
   & &
       \dot \kappa
       + 2 H \kappa
       - 4 \pi G \delta \mu
       - {1 \over 3} \kappa^2
       + {1 \over a^2} \chi^{,i} \kappa_{,i}
       - {1 \over a^4} \left[
       \chi^{,ij} \chi_{,ij}
       - {1 \over 3} \left( \Delta \chi \right)^2 \right]
       =
       {2 \varphi \chi^{,i} \kappa_{,i} \over a^2 (1 + 2 \varphi)}
       - {4 \varphi ( 1 + \varphi ) \over a^4 (1 + 2 \varphi)^2}
       \left[ \chi^{,ij} \chi_{,ij}
       - {1 \over 3} \left( \Delta \chi \right)^2 \right]
   \nonumber \\
   & & \qquad
       + {2 \over a^4 (1 + 2 \varphi)^3}
       \left\{
       {2 \over 3} \left( \Delta \chi \right) \chi^{,i} \varphi_{,i}
       - 2 \chi^{,ij} \chi_{,i} \varphi_{,j}
       + {1 \over 1 + 2 \varphi} \left[
       {1 \over 3} \left( \chi^{,i} \varphi_{,i} \right)^2
       + \chi^{,i} \chi_{,i} \varphi^{,j} \varphi_{,j} \right] \right\},
   \label{eq4-CG-mde-pure-GR} \\
   & & \left( \kappa
       + {\Delta \over a^2} \chi \right)_{,i}
       =
       {2 \varphi \Delta \chi_{,i} \over a^2 ( 1 + 2 \varphi)}
   \nonumber \\
   & & \qquad
       + {1 \over a^2 ( 1 + 2 \varphi)^2}
       \left[
       2 \left( \Delta \chi \right) \varphi_{,i}
       + {1 \over 2} \chi^{,k} \varphi_{,ik}
       - \chi_{,ik} \varphi^{,k}
       + {3 \over 2} \chi_{,i} \Delta \varphi
       - {3 \over 2} {1 \over 1 + 2 \varphi}
       \left( \chi_{,i} \varphi_{,k}
       + {1 \over 3} \chi_{,k} \varphi_{,i} \right) \varphi^{,k} \right].
   \label{eq3-CG-mde-pure-GR}
\eea Terms in the right-hand-sides are pure Einstein's gravity corrections; see the next Section. In the zero-pressure case Newtonian perturbation is closed at the second order in perturbations (Peebles 1980; Vishniac 1983; Zel'dovich \& Novikov 1983). Notice that the pure Einstein's gravity contributions involve $\varphi$ which is related to the (spatial) curvature perturbation in the comoving gauge; see equation (\ref{intrinsic-curvature}).
\end{widetext}

\subsection{Relativistic/Newtonian correspondence}
                                             \label{sec:correspondence}

Except for terms involving $\varphi$ equations
(\ref{eq6-CG-mde-pure-GR})-(\ref{eq3-CG-mde-pure-GR}) coincide
exactly with the Newtonian hydrodynamic equations of the mass and
the momentum conservation equations (removing the gravitational
potential in the momentum conservation equation using the Poisson's
equation), respectively. This statement is true to the fully
nonlinear order in perturbation in the presence of the cosmological
constant in the background. That is, by ignoring $\varphi$ terms (we
cannot do this in general, though) equations (\ref{eq6-CG-mde}),
(\ref{eq4-CG-mde}) and (\ref{eq3-CG-mde}) give \bea
   & &
       \dot \delta - \kappa = \delta \kappa
       - {1 \over a^2} \chi^{,i} \delta_{,i},
   \label{eq6-CG-mde-2} \\
   & &
       \dot \kappa
       + 2 H \kappa
       - 4 \pi G \left( \delta \mu + 3 \delta p \right)
   \nonumber \\
   & & \qquad
       = {1 \over 3} \kappa^2
       - {1 \over a^2} \chi^{,i} \kappa_{,i}
       + {1 \over a^4} \left[
       \chi^{,ij} \chi_{,ij}
       - {1 \over 3} \left( \Delta \chi \right)^2 \right],
   \label{eq4-CG-mde-2} \\
   & &
       \kappa + {\Delta \over a^2} \chi = 0.
   \label{eq3-CG-mde-2}
\eea
By identifying $\delta$ and ${\bf u}$ as the Newtonian density and velocity perturbations with \bea
   & & \kappa \equiv - {1 \over a} \nabla \cdot {\bf u},
\eea thus $\chi = a u$ with ${\bf u} \equiv \nabla u$, we have \bea
   & & \dot \delta + {1 \over a} \nabla \cdot {\bf u}
       = - {1 \over a} \nabla \cdot \left( \delta {\bf u} \right),
   \\
   & & {1 \over a} \nabla \cdot \left( \dot {\bf u} + H {\bf u} \right)
       + 4 \pi G \varrho \delta
       = - {1 \over a^2} \nabla \cdot \left( {\bf u} \cdot \nabla {\bf u} \right),
\eea which coincide exactly with the continuity equation and the Euler equation with the gravitational potential removed using the Poisson's equation (Peebles 1980; Vishniac 1983; Noh \& Hwang 2004).

We emphasize that the above coincidence between Newtonian theory and
Einstein's gravity in the absence of $\varphi$ does not imply that
we can ignore the $\varphi$ terms in any sense. Rather it implies
that in the nonlinear perturbation theory, the pure Einstein's
gravity effect appears purely through the presence of $\varphi$
terms in various ways as in equations
(\ref{eq6-CG-mde-pure-GR})-(\ref{eq3-CG-mde-pure-GR}), and it starts
to appear from the third order perturbation. Thus, to the second
order perturbation we do have exact relativistic/Newtonian
correspondences of the density and velocity perturbations in the
comoving gauge (Noh \& Hwang 2004).

In the comoving gauge, however we do {\it not} have the Newtonian
gravitational potential. In the conventional Newtonian limit of
Einstein's gravity, $\alpha$ usually corresponds to Newtonian
gravitational potential. As in equation (\ref{alpha-CG}), $\alpha$
vanishes to the linear order, and we no longer have the proper
Newtonian correspondence for the gravitational potential in the
comoving gauge. In this regards, the correspondence of density and
velocity perturbation in the two theories to the second order can be
regarded as a coincidence. The proper Newtonian limit can be
achieved in the infinite-speed-of-light limit (this implies the
weak-gravity, slow-motion, negligible pressure, and subhorizon scale
limits) in the zero-shear gauge and the uniform-expansion gauge
(Chandrasekhar 1995; Kofman \& Pogosyan 1995). The proof in our
present formulation is presented separately in Hwang \& Noh (2013).

\section{Other fundamental gauges}
                                             \label{sec:other-gauges}

\subsection{Zero-shear gauge}
                                             \label{sec:ZSG}

In the zero-shear gauge we simply set \bea
   & & \chi \equiv 0.
\eea
We have $N_i = 0 = \overline{K}^i_j$. The metric becomes \bea
   & & ds^2 = - a^2 \left( 1 + 2 \alpha \right) d \eta^2
       + a^2 \left( 1 + 2 \varphi \right) \delta_{ij} d x^i d x^j.
\eea In the linear theory, the zero-shear gauge is quite popular in
the literature (Mukhanov et al. 1992), despite its shortcomings in
numerical treatment in the early universe and in the Boltzmann code
(Ma \& Bertschinger 1995; Hwang \& Noh 2001). Equations
(\ref{eq1})-(\ref{eq7}) give a (redundantly) complete set of
equations for the nonlinear perturbation variables $\delta$, $v$,
$\kappa$, $\varphi$ and $\alpha$. All perturbation variables in the
zero-shear gauge are gauge invariant with \bea
   & & \delta = \delta_\chi, \quad
      v = v_\chi, \quad
      \kappa = \kappa_\chi, \quad
      \varphi = \varphi_\chi, \quad
      \alpha = \alpha_\chi.
\eea There are several ways of having the closed form second order
differential equations. Equations (\ref{eq6}) and (\ref{eq4}),
together with equations (\ref{eq2}), (\ref{eq3}) and (\ref{eq5}) to
determine $\alpha$, $\varphi$ and $\chi$, give equations for $\dot
\delta$ and $\dot \kappa$. Equations (\ref{eq6}) and
(\ref{eq7}), together with equations (\ref{eq5}), (\ref{eq2}) and
(\ref{eq3}) to determine $\alpha$, $\varphi$ and $\kappa$, give
equations for $\dot \delta$ and $\dot v$.
Etc.

It is known that in the small-scale limit (i.e., inside the visual
horizon) in the matter dominated era, the density, velocity and
gravitational potential variables ($\delta$, $v$ and $\alpha$) in
the zero-shear gauge coincide exactly with the Newtonian results
(Hwang \& Noh 1999); this is true even for the variables in the
uniform-expansion gauge. We have shown that the correspondences
continue to be valid even to the second-order perturbation (Hwang et al. 2012) and to fully non-linear and exact order in the infinite-speed-of-light limit (Hwang \& Noh 2013).

\subsection{Uniform-curvature gauge}
                                             \label{sec:UCG}

In the uniform-curvature gauge we set \bea
   & & \varphi \equiv 0.
\eea
We have $R^{(h)i}_{\;\;\;\;\;\; jk\ell} = 0$, thus flat in the flat background. The metric becomes \bea
   & & ds^2 = - a^2 \left( 1 + 2 \alpha \right) d \eta^2
       - 2 a \chi_{,i} d \eta d x^i
       + a^2 \delta_{ij} d x^i d x^j.
   \nonumber \\
\eea
Equations (\ref{eq1})-(\ref{eq7}) give a complete set of equations for variables $\delta$, $v$, $\kappa$, $\chi$ and $\alpha$. All perturbation variables in the uniform-curvature gauge are gauge invariant with \bea
   & & \delta = \delta_\varphi, \quad
      v = v_\varphi, \quad
      \kappa = \kappa_\varphi, \quad
      \chi = \chi_\varphi, \quad
      \alpha = \alpha_\varphi.
\eea In the linear theory, the uniform-curvature gauge is useful to
handle the scalar field perturbation (Field \& Shepley 1968; Lukash
1980a, 1980b; Sasaki 1986; Mukhanov 1988; Hwang 1994; Hwang \& Noh
2005). To the linear order, with the scalar field $\widetilde \phi$
decomposed as $\widetilde \phi = \phi + \delta \phi$, we have
$\delta \widehat \phi = \delta \phi - \phi^\prime \xi^0$, thus \bea
   & & \delta \phi_\varphi \equiv \delta \phi
       - {\dot \phi \over H} \varphi
       \equiv - {\dot \phi \over H} \varphi_{\delta \phi},
\eea is gauge invariant. For fully nonlinear treatment of the scalar field perturbation in the uniform-field gauge ($\delta \phi \equiv 0$ to the nonlinear order), see Section \ref{sec:MSF-CG}.

\subsection{Uniform-expansion gauge}
                                             \label{sec:UEG}

In the uniform-expansion gauge we set \bea
   & & \kappa \equiv 0.
\eea
We have $\widetilde \theta^{(n)} = - K = 3 H$, thus the expansion rate of the normal frame vector field is uniform in space. Equations (\ref{eq1})-(\ref{eq7}) give a complete set of equations for variables  $\delta$, $v$, $\chi$, $\varphi$ and $\alpha$. All perturbation variables in the uniform-expansion gauge are gauge invariant with \bea
   & & \delta = \delta_\kappa, \quad
      v = v_\kappa, \quad
      \varphi = \varphi_\kappa, \quad
      \chi = \chi_\kappa, \quad
      \alpha = \alpha_\kappa.
\eea As in the zero-shear gauge, the uniform-expansion gauge also
shows small-scale Newtonian correspondence of $\delta$, $v$ and
$\alpha$ up to the second order in perturbation (Hwang et al. 2012). It also has correct Newtonian limit in the infinite-speed-of-light limit (Hwang \& Noh 2013).

\subsection{Uniform-density gauge}
                                             \label{sec:UDG}

In the uniform-density gauge we set \bea
   & & \delta \equiv 0,
\eea thus the density is uniform in the hypersurface.
Equations (\ref{eq1})-(\ref{eq7}) give a complete set of equations for variables $v$, $\kappa$, $\chi$, $\varphi$ and $\alpha$. All perturbation variables in the uniform-density gauge are gauge invariant with \bea
   & & v = v_\delta, \quad
      \kappa = \kappa_\delta, \quad
      \varphi = \varphi_\delta, \quad
      \chi = \chi_\delta, \quad
      \alpha = \alpha_\delta.
\eea

The curvature perturbation $\varphi$ in the comoving gauge
($\varphi_v$), in the uniform expansion gauge ($\varphi_\kappa$) and
in the uniform-density guage ($\varphi_\delta$) are known to have nice
conservation behaviors in the large scale (super-sound-horizon scale) limit to the second order in perturbations (Hwang \& Noh 2007), and to general
nonlinear order based on the spatial gradient expansion method (Lyth
et al. 2005). The proof based on our exact and fully nonlinear
perturbation equations deserves a further examination.

\section{Generation of vorticity}
                                             \label{sec:rotation}

We consider the generation of vorticity (rotation or vector-type
perturbation) from the pure scalar-type perturbation. We consider
the comoving gauge. Thus, we have $v_i = 0$ in the $v_i$ terms
multiplied by perturbation terms, and have $v_i = v^{(v)}_i$ in the
pure $v_i$ term without perturbations multiplied. From equation
(\ref{eq7}) we have \bea
   & & {1 \over a^3} \left[ a^4
       \left( \mu + p \right) v^{(v)}_i \right]^{\displaystyle\cdot}
       = - \widetilde p_{,i}
       - \left( \widetilde \mu + \widetilde p \right)
       {{\cal N}_{,i} \over {\cal N}}.
\eea Thus, we have \bea
   & & {1 \over a^4} \left[ a^4
       \left( \mu + p \right) v^{(v)}_{[i,j]} \right]^{\displaystyle\cdot}
       = - { \widetilde \mu_{,[i} \widetilde p_{,j]}
       \over a (\mu + p)},
   \label{vorticity-generation-eq}
\eea which is true to the fully nonlinear order; notice that the
right-hand-side vanishes for an ideal fluid with $\widetilde p =
\widetilde p (\widetilde  \mu)$.

Now, we consider the general case without taking the gauge
condition. In terms of the covariant equations, from equation (8) in
Hawking (1966) we have \bea
   & & \widetilde h^c_a \widetilde h^d_b \left(
       \widetilde {\dot {\widetilde \omega}}_{cd}
       - \widetilde a_{[c;d]} \right)
       = - {2 \over 3} \widetilde \theta \widetilde \omega_{ab}
       + 2 \widetilde \sigma^c_{\;\;[a} \widetilde \omega_{b]c}.
   \label{covariant-rotation-eq}
\eea For the covariant
notations see the Appendix C. Using equation (\ref{eq7-cov}),
equation (\ref{omega-2}) becomes \bea
   & & \widetilde \omega_{ij}
       = a v^{(v)}_{[i,j]}
       + {1 \over \widetilde \mu + \widetilde p}
       a v_{[i} \widetilde p_{,j]}.
   \label{vorticity-pressure}
\eea For the vorticity generation from the pure scalar-type
perturbation in the comoving gauge, we have $\widetilde \omega_{ij}
= a v^{(v)}_{[i,j]}$, and from equations
(\ref{covariant-rotation-eq}), (\ref{covariant-E-conserv}) and
(\ref{covariant-mom-conserv}) we arrive at equation
(\ref{vorticity-generation-eq}).


The examination of equation (\ref{vorticity-pressure}) leads to the
following conclusions valid to the fully nonlinear order.
In the absence of pressure, we have $\widetilde
\omega_{ij} = 0$ for $v_i^{(v)} = 0$ independently of the gauge
condition (we wish to thank the referee for clarifying comments on our previously confused state).
This conclusion is consistent with equation
(\ref{covariant-rotation-eq}) which shows that in the absence of
pressure equation of $\widetilde \omega_{ab}$ becomes homogeneous
[we have $\widetilde a_c = 0$ from equation
(\ref{covariant-mom-conserv})], thus an irrotational ($\widetilde
\omega_{ab} = 0$) fluid remains irrotational independently of the
gauge condition.
For works on related issues, see Christopherson et al (2009) and Lu et al (2009).

\section{Generation of gravitational waves from scalar- and vector-type perturbations}
                                             \label{sec:GW}

Fully nonlinear and exact formulation including the tensor-type perturbation is supposed to be a complicated subject which is left for future investigation. Here we consider a much simpler case with linear tensor-type perturbation. In the presence of the tensor-type perturbation a change occurs in the spatial part of the metric in equation (\ref{metric-convention}) as \bea
   & & \widetilde g_{ij}
       = a^2 \left[ \left( 1 + 2 \varphi \right) \delta_{ij}
       + 2 h_{ij} \right],
\eea where $h_{ij}$ is the transverse ($h^j_{i|j} \equiv 0$) and tracefree ($h^i_i \equiv 0$) tensor-type perturbation; indices of $h_{ij}$ are raised and lowered by $\delta_{ij}$ as the metric; only in this section $h_{ij}$ indicates the tensor-type perturbation. By keeping only linear order terms in $h_{ij}$ we can update quantities in the Appendix B. In our basic perturbation equations in (\ref{eq1})-(\ref{eq7}), the linear tensor-type perturbation contributes {\it only} in equation (\ref{eq5}) by simply adding the following term \bea
   & & \ddot h^i_j
       + 3 H \dot h^i_j
       - {\Delta \over a^2} h^i_j,
\eea in the left-hand-side. We ignore the tensor-type anisotropic stress. Let us write equation (\ref{eq5}) including the linear tensor-type perturbation as \bea
   & & \ddot h_{ij}
       + 3 H \dot h_{ij}
       - {\Delta \over a^2} h_{ij} \equiv n_{ij},
\eea where we moved all the terms in equation (\ref{eq5}) to the right-hand-side and called it $n^i_j$, then lowered the index by $\delta_{ij}$. The right-hand-side of this equation includes linear parts of the scalar- and vector-type perturbations which need to be removed to get the pure tensor-type perturbation equation generated by the nonlinear scalar- and vector-type perturbations.  By the following operation we can separate the linear part of the scalar- and vector-type contributions [see equation (210) in Noh \& Hwang (2004)] \bea
   & & \ddot h_{ij}
       + 3 H \dot h_{ij}
       - {\Delta \over a^2} h_{ij} = s_{ij},
   \\
   & & s_{ij} \equiv n_{ij}
       - 2 \Delta^{-1} \nabla_{(i} n^k_{j),k}
   \nonumber \\
   & & \qquad
       + {1 \over 2} \Delta^{-2}
       \left( \nabla_i \nabla_j + \delta_{ij} \Delta \right) n^{kl}_{\;\;\;|kl}.
\eea
This can be regarded as the equation for gravitational waves (tensor-type perturbation) generated from pure scalar- and vector-type perturbations to the fully nonlinear order. The nonlinear terms in $s_{ij}$ still depend on the temporal gauge condition, and consequently the gravitational waves generated from the scalar- and vector-type perturbations do depend on the temporal gauge choice. In the perturbation approach $h_{ij}$ should be regarded as the same order perturbation as the one considered in $s_{ij}$; i.e., with an expansion $h_{ij} = h_{ij}^{(1)} + h_{ij}^{(2)} + \dots$, where index $(1)$ and $(2)$ indicating the order of perturbation, for $s_{ij}$ quadratic order perturbations $h_{ij}$ is the same as $h_{ij}^{(2)}$, etc., thus depending on the gauge choice. This subject deserves further studies.

\section{Scalar field in the comoving gauge}
                                             \label{sec:MSF-CG}

We consider a minimally coupled scalar field $\widetilde \phi$ with \bea
   & & \widetilde T_{ab} = \widetilde \phi_{,a} \widetilde \phi_{,b}
       - \left[ {1 \over 2} \widetilde \phi^{,c} \widetilde \phi_{,c} + \widetilde V (\widetilde \phi) \right] \widetilde g_{ab},
   \label{Tab-MSF}
\eea and the equation of motion \bea
   & & \widetilde \phi^{;c}_{\;\;\; c} = \widetilde V_{,\widetilde \phi}.
   \label{EOM-MSF}
\eea The fluid quantities can be read from equation (\ref{Tab-decomposition}).

The fluid quantities and the equation of motion to the fully nonlinear order can be derived in a gauge-ready form. Here, we only consider a fluid formulation of the scalar field in the comoving gauge; the gauge-ready formulation is presented in the Appendix \ref{sec:MSF}. From equations (\ref{Tab}) and (\ref{Tab-MSF}) we have \bea
   & & \widetilde u_i = - {1 \over \widetilde \mu} \widetilde T_{ib} \widetilde u^b
       = - {\widetilde \phi_{,i} \over \widetilde \phi_{,c} \widetilde u^c}.
\eea Thus, the comoving gauge ($v \equiv 0$) implies the uniform-field gauge ($\delta \phi \equiv 0$) and {\it vice versa} to the fully nonlinear order; we set $\widetilde \phi = \phi + \delta \phi$ where $\phi$ is the background order scalar field. In this gauge we can show \bea
   & & \widetilde \mu = {1 \over 2 {\cal N}^2} \dot \phi^2
       + V, \quad
       \widetilde p = {1 \over 2 {\cal N}^2} \dot \phi^2
       - V, \quad
       \widetilde \pi_{ab} = 0.
\eea Thus, to the background order, we have \bea
   & & \mu = {1 \over 2} \dot \phi^2 + V, \quad
       p = {1 \over 2} \dot \phi^2 - V,
\eea and to the fully nonlinear order, we have \bea
   & & \delta p = \delta \mu = - {1 \over 2 {\cal N}^2} \dot \phi^2
       \left( 2 \alpha
       + {\chi^{k} \chi_{k} \over a^2 ( 1 + 2 \varphi )} \right),
\eea with vanishing anisotropic stress. Therefore, the ideal fluid equations in equations (\ref{eq1})-(\ref{eq7}) under the comoving gauge remain valid with the perturbed equation of state given as $\delta p = \delta \mu$.

In the Appendix \ref{sec:MSF} we present the full formulation in the case of a minimally coupled scalar field.

\section{Discussion}
                                                     \label{sec:Discussion}

Extension or feasibility of similar fully nonlinear formulation including the following cases deserves future investigations:
(i) background spatial curvature,
(ii) anisotropic stress,
(iii) scalar fields (see the Appendix \ref{sec:MSF} for a minimally coupled scalar field),
(iv) multiple components of fluids and fields,
(v) class of generalized gravity theories,
(vi) the electric and magnetic fields,
(vii) the covariant equations and the Weyl tensors,
(viii) the tensor-type perturbation,
(ix) null geodesic for Sachs-Wolfe effect,
(x) gravitational lensing,
(xi) Boltzmann equations for photons, and massless and massive neutrinos,
(xii) gauge transformation properties,
(xiii) expression of gauge-invariant combinations,
(xiv) equations in mixed gauge conditions,
etc.
At the moment the first seven are trivial (some could be tedious though) extensions while the remaining ones need closer examinations for their feasibilities. Implementations of all the above cases (except for x) were made to the second order in perturbations in Noh \& Hwang (2004), Hwang \& Noh (2007), and Hwang et al (2012).

Neglecting the transverse-tracefree (TT) part of the metric assumed
in equation (\ref{metric-convention}) is indeed a quite serious
constraint and drawback on our formulation aiming for fully
nonlinear and exact analysis. The referee has point out that ``in general in a non-linear context
the TT part of the metric doesn't necessarily represents
gravitational waves, but more general TT distortions of the spatial
curvature'' (see also Matarrese \& Terranova 1996). The referee has also informed us that
``there are well known stationary spacetimes (therefore not
containing gravitational waves) where the spatial 3-metric cannot be
made conformally flat, for instance Kerr or the metric for a
rotating star.''. At the moment, it looks, the presence of TT part
does not allow us to get the inverse metric in exact form, thus
forbidding us to proceed. As mentioned in section \ref{sec:GW} the
TT part can always be handled perturbatively. Although to the
nonlinear order the TT part is gauge dependent, as we have shown in
section \ref{sec:convention} it does not affect the gauge issue of
the scalar- and vector-type perturbations addressed in this work,
see the nonlinear gauge transformation issue addressed below
equation (\ref{GI}).

We anticipate potentially wide applications of our exact and fully nonlinear perturbation formulation, not only in higher order perturbation theory, but also in the averaging, fitting and back-reaction approaches in theoretical cosmology (Ellis 1984; Ellis \& Stoeger 1987; Clarkson et al 2011).

\section*{Acknowledgments}

We wish to thank Professor Hee-Won Lee for hospitality during J.H.'s sabbatical visit Sejong University. H.N.\ was supported by grant No.\ 2012 R1A1A2038497 from NRF. J.H.\ was supported by KRF Grant funded by the Korean Government
(KRF-2008-341-C00022).

\begin{widetext}

\appendix

\section{ADM ($3+1$) equations review}
                                                     \label{sec:ADM}

The ADM (Arnowitt-Deser-Misner) formulation (Arnowitt et al. 1962)
is based on splitting the spacetime into the spatial and the temporal parts
using a normal four-vector field $\widetilde n_a$.
The metric is written as
\bea
   & & \widetilde g_{00} \equiv - N^2 + N^i N_i, \quad
       \widetilde g_{0i} \equiv N_i, \quad
       \widetilde g_{ij} \equiv h_{ij}, \quad
       \widetilde g^{00} = - N^{-2}, \quad
       \widetilde g^{0i} = N^{-2} N^i, \quad
       \widetilde g^{ij} = h^{ij} - N^{-2} N^i N^j,
   \label{ADM-metric-def}
\eea
where the index of $N_i$ is raised and lowered by $h_{ij}$ as the metric,
and $h^{ij}$ is an inverse metric of $h_{ij}$; for meanings of the ADM variables, see Smarr \& York (1978).
The normal four-vector $\widetilde n_a$ is introduced as
\bea
   & & \widetilde n_0 = - N, \quad
       \widetilde n_i \equiv 0, \quad
       \widetilde n^0 = N^{-1}, \quad
       \widetilde n^i = - N^{-1} N^i.
   \label{n_a-def}
\eea
The fluid quantities are defined as
\bea
   & & E \equiv \widetilde n_a \widetilde n_b \widetilde T^{ab}, \quad
       J_i \equiv - \widetilde n_b \widetilde T^b_i, \quad
       S_{ij} \equiv \widetilde T_{ij}, \quad
       S \equiv h^{ij} S_{ij}, \quad
       \overline{S}_{ij} \equiv S_{ij}
       - {1\over 3} h_{ij} S,
   \label{ADM-fluid-def}
\eea
where the indices of $J_i$ and $S_{ij}$ are raised and lowered by $h_{ij}$.
The extrinsic curvature is introduced as
\bea
   & & K_{ij} \equiv {1\over 2N} \left( N_{i:j}
       + N_{j:i} - h_{ij,0} \right), \quad
       K \equiv h^{ij} K_{ij}, \quad
       \overline{K}_{ij} \equiv K_{ij}
       - {1\over 3} h_{ij} K,
   \label{extrinsic-curvature-def}
\eea
where the index of $K_{ij}$ is raised and lowered by $h_{ij}$.
A colon `$:$' denotes a covariant derivative based on $h_{ij}$.
$\Gamma^{(h)i}_{\;\;\;\;jk}$ is the connection based on
$h_{ij}$ as the metric,
$\Gamma^{(h)i}_{\;\;\;\;\;jk} \equiv {1\over 2} h^{i\ell}
\left( h_{j\ell,k} + h_{k\ell,j} - h_{jk,\ell} \right)$.
The intrinsic curvatures are based on $h_{ij}$ as the metric
\bea
   & & R^{(h)i}_{\;\;\;\;\;\;\;jk\ell}
       \equiv
       \Gamma^{(h)i}_{\;\;\;\;\;j\ell,k}
       - \Gamma^{(h)i}_{\;\;\;\;\;jk,\ell}
       + \Gamma^{(h)m}_{\;\;\;\;\;j\ell}
       \Gamma^{(h)i}_{\;\;\;\;\;km}
       - \Gamma^{(h)m}_{\;\;\;\;\;jk}
       \Gamma^{(h)i}_{\;\;\;\;\;\ell m},
   \nonumber \\
   & &
       R^{(h)}_{ij}
       \equiv R^{(h)k}_{\;\;\;\;\;\;\;ikj}, \quad
       R^{(h)} \equiv h^{ij} R^{(h)}_{ij}, \quad
       \overline{R}^{(h)}_{ij} \equiv R^{(h)}_{ij}
       - {1\over 3} h_{ij} R^{(h)}.
   \label{ADM-curvature}
\eea
A complete set of the ADM equations is the following (Bardeen 1980; Noh \& Hwang 2004)
\bea
   & & R^{(h)} = \overline{K}^{ij} \overline{K}_{ij}
       - {2 \over 3} K^2 + 16 \pi G E + 2 \Lambda,
   \label{E-constraint} \\
   & & \overline{K}^j_{i:j} - {2 \over 3} K_{,i}
       = 8 \pi G J_i,
   \label{Mom-constraint} \\
   & & K_{,0} N^{-1} - K_{,i} N^i N^{-1}
       + N^{:i}_{\;\;\;i} N^{-1}
       - \overline{K}^{ij} \overline{K}_{ij}
       - {1\over 3} K^2 - 4 \pi G \left( E + S \right) + \Lambda = 0,
   \label{Trace-prop} \\
   & & \overline{K}^i_{j,0} N^{-1}
       - \overline{K}^i_{j:k} N^k N^{-1}
       + \overline{K}_j^k N^i_{\;\;:k} N^{-1}
       - \overline{K}^i_k N^k_{\;\;:j} N^{-1}
       = K \overline{K}^i_j
       - \left( N^{:i}_{\;\;\;j}
       - {1\over 3} \delta^i_j N^{:k}_{\;\;\;k} \right)
       N^{-1}
       + \overline{R}^{(h)i}_{\;\;\;\;\;j} - 8 \pi G \overline{S}^i_j,
   \label{Tracefree-prop} \\
   & & E_{,0} N^{-1} - E_{,i} N^i N^{-1}
       - K \left( E + {1 \over 3} S \right)
       - \overline{S}^{ij} \overline{K}_{ij}
       + N^{-2} \left( N^2 J^i \right)_{:i}
       = 0,
   \label{E-conservation} \\
   & & J_{i,0} N^{-1} - J_{i:j} N^j N^{-1}
       - J_j N^j_{\;\;:i} N^{-1} - K J_i
       + E N_{,i} N^{-1}
       + S^j_{i:j}
       + S_i^j N_{,j} N^{-1} = 0.
   \label{Mom-conservation}
\eea These are the ADM energy constraint, the ADM momentum constraint, trace of ADM propagation, tracefree ADM propagation, the ADM energy conservation, and the ADM momentum conservation equations, respectively. Equations (\ref{E-constraint})-(\ref{Mom-conservation}) together with definition of $K$ in equation (\ref{extrinsic-curvature-def}) provide the fundamental perturbation equations presented in equations (\ref{eq1})-(\ref{eq7}). In the Appendix B we present details of steps useful for the derivation.

\section{Derivation of fully nonlinear perturbations}
                                          \label{sec:derivation}

Our metric convention is \bea
   & & \widetilde g_{00} = - a^2 \left( 1 + 2 \alpha \right), \quad
       \widetilde g_{0i} = - a \chi_{i}, \quad
       \widetilde g_{ij} = a^2 \left( 1 + 2 \varphi \right) \delta_{ij},
   \label{metric-A}
\eea where the index of $\chi_i$ is raised and lowered by $\delta_{ij}$ as the metric. We {\it assume} $a$ is a function of conformal time ($x^0 = \eta$) only, whereas $\alpha$, $\chi_i$ and $\varphi$ are general functions of space and time, but we do {\it not} assume these to be small in amplitudes. Justification of our metric convention is made in Section \ref{sec:convention}.
The inverse metric is \bea
   & & \widetilde g^{00}
       = - {1 \over a^2} {1 + 2 \varphi \over (1 + 2 \varphi) (1 + 2 \alpha)
       + \chi^{k} \chi_{k} / a^2}, \quad
       \widetilde g^{0i} = - {1 \over a^2} {\chi^{i}/a \over (1 + 2 \varphi) (1 + 2 \alpha)
       + \chi^{k} \chi_{k} / a^2},
   \nonumber \\
   & &
       \widetilde g^{ij} = {1 \over a^2 ( 1 + 2 \varphi)}
       \left( \delta^{ij} - { \chi^{i} \chi^{j} / a^2 \over
       (1 + 2 \varphi) (1 + 2 \alpha)
       + \chi^{k} \chi_{k} / a^2} \right).
\eea
From equation (\ref{ADM-metric-def}) the ADM metric can be identified as \bea
   & & N = a \sqrt{ 1 + 2 \alpha + {\chi^{k} \chi_{k} \over
       a^2 (1 + 2 \varphi)}}
       \equiv a {\cal N}, \quad
       N_i = - a \chi_{i}, \quad
       N^i = - {\chi^{i} \over a (1 + 2 \varphi)},
   \nonumber \\
   & &
       h_{ij} = a^2 \left( 1 + 2 \varphi \right) \delta_{ij}, \quad
       h^{ij} = {1 \over a^2 (1 + 2 \varphi)} \delta^{ij},
   \label{ADM-metric}
\eea thus  \bea
   & & \widetilde g^{00}
       = - {1 \over a^2 {\cal N}^2}, \quad
       \widetilde g^{0i} = - {\chi^{i} \over a^3 {\cal N}^2 (1 + 2 \varphi)},
       \quad
       \widetilde g^{ij} = {1 \over a^2 ( 1 + 2 \varphi)}
       \left( \delta^{ij} - { \chi^{i} \chi^{j} \over a^2 {\cal N}^2
       (1 + 2 \varphi)} \right),
   \label{metric-A-inverse}
\eea and
\bea
   & & \widetilde n_i \equiv 0, \quad
       \widetilde n_0 = - a {\cal N}, \quad
       \widetilde n^i = {\chi^i \over a^2 {\cal N} ( 1 + 2 \varphi )}, \quad
       \widetilde n^0 = {1 \over a {\cal N}}.
   \label{n_a}
\eea
The three-space connection and curvatures are \bea
   & & \Gamma^{(h)i}_{\;\;\;\;\;jk}
       = {1 \over 1 + 2 \varphi} \left(
       \varphi_{,j} \delta^i_k
       + \varphi_{,k} \delta^i_j
       - \varphi^{,i} \delta_{jk} \right), \quad
       \Gamma^{(h)k}_{\;\;\;\;\;ik}
       = {3 \varphi_{,i} \over 1 + 2 \varphi},
   \nonumber \\
   & & R^{(h)}_{ij}
       = - {\varphi_{,ij} \over 1 + 2 \varphi}
       + 3 {\varphi_{,i} \varphi_{,j} \over (1 + 2 \varphi)^2}
       - \left( {\Delta \varphi \over 1 + 2 \varphi}
       - {\varphi^{,k} \varphi_{,k} \over (1 + 2 \varphi)^2} \right) \delta_{ij}, \quad R^{(h)} = {2 \over a^2 (1 + 2 \varphi)^2}
       \left( - 2 \Delta \varphi
       + 3 {\varphi^{,k} \varphi_{,k} \over 1 + 2 \varphi} \right),
   \nonumber \\
   & & \overline{R}^{(h)i}_{\;\;\;\;\;j}
       = {1 \over a^2 (1 + 2 \varphi)^2} \left[
       - \varphi^{,i}_{\;\;j}
       + 3 {\varphi^{,i} \varphi_{,j} \over 1 + 2 \varphi}
       - {1 \over 3} \delta^i_j \left( - \Delta \varphi
       + 3 {\varphi^{,k} \varphi_{,k} \over 1 + 2 \varphi} \right) \right].
   \label{intrinsic-curvature}
\eea
The extrinsic curvature gives \bea
   & & K_{ij}
       = - {a^2 \over {\cal N}} \left[ \left( H + \dot \varphi + 2 H \varphi \right) \delta_{ij}
       + {1 \over 2 a^2} \left( \chi_{i,j} + \chi_{j,i} \right)
       - {1 \over a^2 (1 + 2 \varphi)} \left(
       \chi_{i} \varphi_{,j}
       + \chi_{j} \varphi_{,i}
       - \chi^{k} \varphi_{,k} \delta_{ij} \right) \right],
   \nonumber \\
   & & K
       = - {1 \over {\cal N} (1 + 2 \varphi)}
       \left[ 3 \left( H + \dot \varphi + 2 H \varphi \right)
       + {1 \over a^2} \chi^k_{\;\;,k}
       + {\chi^{k} \varphi_{,k} \over a^2 (1 + 2 \varphi)}
       \right]
       \equiv - 3 H + \kappa,
   \nonumber \\
   & & \overline{K}^i_j
       = - {1 \over a^2 {\cal N} (1 + 2 \varphi)} \left[
       {1 \over 2} \left( \chi^{i}_{\;\;,j} + \chi_j^{\;,i} \right)
       - {1 \over 3} \delta^i_j \chi^k_{\;\;,k}
       - {1 \over 1 + 2 \varphi} \left(
       \chi^{i} \varphi_{,j}
       + \chi_{j} \varphi^{,i}
       - {2 \over 3} \delta^i_j \chi^{k} \varphi_{,k} \right)
       \right],
   \nonumber \\
   & & \overline{K}^i_j \overline{K}^j_i
       = {1 \over a^4 {\cal N}^2 (1 + 2 \varphi)^2}
       \Bigg\{
       {1 \over 2} \chi^{i,j} \left( \chi_{i,j} + \chi_{j,i} \right)
       - {1 \over 3} \chi^i_{\;\;,i} \chi^j_{\;\;,j}
       - {4 \over 1 + 2 \varphi} \left[
       {1 \over 2} \chi^i \varphi^{,j} \left(
       \chi_{i,j} + \chi_{j,i} \right)
       - {1 \over 3} \chi^i_{\;\;,i} \chi^j \varphi_{,j} \right]
   \nonumber \\
   & & \qquad
       + {2 \over (1 + 2 \varphi)^2} \left(
       \chi^{i} \chi_{i} \varphi^{,j} \varphi_{,j}
       + {1 \over 3} \chi^i \chi^j \varphi_{,i} \varphi_{,j} \right) \Bigg\}.
   \label{extrinsic-curvature}
\eea
The fluid four-vector is \bea
   & & \widetilde u_i \equiv a v_{i}, \quad
       \widetilde u_0 =
       - a {\cal N} \sqrt{ 1 + {v^{k} v_{k} \over 1 + 2 \varphi} }
       - {\chi^{k} v_{k} \over 1 + 2 \varphi},
   \nonumber \\
   & & \widetilde u^i = {v^{i} \over a (1 + 2 \varphi)}
       + {\chi^{i} \over a^2 {\cal N} (1 + 2 \varphi)}
       \sqrt{ 1 + {v^{k} v_{k} \over 1 + 2 \varphi} }, \quad
       \widetilde u^0
       = {1 \over a {\cal N}} \sqrt{ 1 + {v^{k} v_{k}
       \over 1 + 2 \varphi} },
   \label{u_a}
\eea where the index of $v_i$ is raised and lowered by $\delta_{ij}$ as the metric. For $v_i = 0$ we have $\widetilde u_a = \widetilde n_a$.
The energy-momentum tensor of an ideal fluid is \bea
   & & \widetilde T^0_0
       = - \widetilde \mu
       - {\widetilde \mu + \widetilde p \over 1 + 2 \varphi}
       \left( v^i v_i
       + {1 \over a {\cal N}} \chi^i v_i
       \sqrt{ 1 + {v^{k} v_{k} \over 1 + 2 \varphi} } \right), \quad
       \widetilde T^0_i
       = {1 \over {\cal N}} \left( \widetilde \mu + \widetilde p \right)
       v_i \sqrt{ 1 + {v^{k} v_{k} \over 1 + 2 \varphi} },
   \nonumber \\
   & &
       \widetilde T_{ij}
       = a^2 \left[ \left( 1 + 2 \varphi \right) \widetilde p \delta_{ij}
       + \left( \widetilde \mu + \widetilde p \right) v_i v_j \right].
   \label{Tab-pert}
\eea
The ADM fluid quantities become \bea
   & & E = \widetilde \mu
       + \left( \widetilde \mu + \widetilde p \right) {v^{k} v_{k} \over 1 + 2 \varphi}, \quad
       J_i = a \left( \widetilde \mu + \widetilde p \right) v_{i}
       \sqrt{ 1 + {v^{k} v_{k} \over 1 + 2 \varphi}}, \quad
       J^i = {\widetilde \mu + \widetilde p \over a (1 + 2 \varphi)} v^{i}
       \sqrt{ 1 + {v^{k} v_{k} \over 1 + 2 \varphi}},
   \nonumber \\
   & & S^i_j
       = \widetilde p \delta^i_j
       + \left( \widetilde \mu + \widetilde p \right)
       {v^{i} v_{j} \over 1 + 2 \varphi}, \quad
       S = 3 \widetilde p + {\widetilde \mu + \widetilde p \over 1 + 2 \varphi}
       v^{k} v_{k}, \quad
       \overline{S}^i_j
       = {\widetilde \mu + \widetilde p \over 1 + 2 \varphi}
       \left( v^{i} v_{j} - {1 \over 3} \delta^i_j v^{k} v_{k} \right).
\eea Notice that for $v_i = 0$ (ignoring the vector-type
perturbation $v^{(v)}_i = 0$, and taking the comoving gauge $v
\equiv 0$) the energy-momentum tensor and the ADM fluid quantities
are simplified to \bea
   & & \widetilde T^0_0 = - \widetilde \mu, \quad
       \widetilde T^0_i = 0, \quad
       \widetilde T_{ij} = a^2 \left( 1 + 2 \varphi \right)
       \widetilde p \delta_{ij}, \quad
       E = \widetilde \mu, \quad
       J_i = 0, \quad
       S = 3 \widetilde p, \quad
       \overline{S}^i_j = 0.
\eea
Using the above quantities, from equations (\ref{extrinsic-curvature-def}) for $K$ and equations (\ref{E-constraint})-(\ref{Mom-conservation}) we can derive equations (\ref{eq1})-(\ref{eq7}), respectively.

\section{Covariant formulation}
                                          \label{sec:cov}

Here we present the covariant kinematic quantities based on the normalized fluid four-vector $\widetilde u_a$. We have (Ehlers 1993; Ellis 1971, 1973)
\bea
   & & \widetilde h^c_a \widetilde h^d_b \widetilde u_{c;d}
       = \widetilde h^c_{[a} \widetilde h^d_{b]} \widetilde u_{c;d}
       + \widetilde h^c_{(a} \widetilde h^d_{b)} \widetilde u_{c;d}
       \equiv \widetilde \omega_{ab} + \widetilde \theta_{ab}
       = \widetilde u_{a;b} + \widetilde a_a \tilde u_b,
   \nonumber \\
   & & \widetilde \sigma_{ab} \equiv \widetilde \theta_{ab}
       - {1 \over 3} \widetilde \theta \widetilde h_{ab}, \quad
       \widetilde \theta \equiv \widetilde u^a_{\;\; ;a}, \quad
       \widetilde a_a \equiv {\widetilde {\dot {\widetilde u}}}_a
       \equiv \widetilde u_{a;b} \widetilde u^b,
   \label{kinematic-shear}
\eea where $\widetilde h_{ab} \equiv \widetilde g_{ab} + \widetilde
u_a \widetilde u_b$ is the projection tensor with $\widetilde h_{ab}
\widetilde u^b = 0$ and $\widetilde h^a_a = 3$; $t_{(ab)} \equiv {1
\over 2} (t_{ab} + t_{ba})$ and $t_{[ab]} \equiv {1 \over 2} (t_{ab}
- t_{ba})$; we have $\widetilde a_c \widetilde u^c = 0$ and
$\widetilde \sigma_{ab} \widetilde u^b = 0 = \widetilde \omega_{ab}
\widetilde u^b$. An overdot with tilde $\widetilde {\dot {}}$
indicates the covariant derivative along $\widetilde u^a$. The
quantities $\widetilde \theta$, $\widetilde a_c$, $\widetilde
\sigma_{ab}$, and $\widetilde \omega_{ab}$ are the expansion scalar,
the acceleration vector, the shear tensor, and the rotation
(vorticity) tensor, respectively, of the $\widetilde u_a$ flow.

Using equations (\ref{ADM-metric-def}), (\ref{ADM-metric}),
(\ref{intrinsic-curvature}) and (\ref{u_a}) we can show [the
spacetime connection $\widetilde \Gamma^a_{bc}$ in terms of the ADM
notations is presented in equation (6) of Noh \& Hwang (2004)] \bea
   & & \widetilde {\dot {\widetilde \mu}}
       = \left[ {1 \over a {\cal N}}
       \sqrt{ 1 + {v^k v_k \over 1 + 2 \varphi} }
       \partial_0
       + {1 \over a (1 + 2 \varphi)} \left( v^\ell
       + \sqrt{ 1 + {v^k v_k \over 1 + 2 \varphi} } {\chi^\ell \over a {\cal N}}
       \right)
       \partial_\ell \right]
       \widetilde \mu,
   \\
   & & \widetilde \theta
       = \left( 3 H - \kappa \right) \sqrt{ 1 + {v^k v_k \over 1 + 2 \varphi} }
       + { ({\cal N} v^i)_{,i} \over a {\cal N} ( 1 + 2 \varphi )}
       + {v^i \varphi_{,i} \over a (1 + 2 \varphi)^2}
       + {1 \over a {\cal N}} \left( \partial_0
       + {\chi^i \over a (1 + 2 \varphi)} \nabla_i \right)
        \sqrt{ 1 + { v^k v_k \over 1 + 2 \varphi} },
   \\
   & & \widetilde a_i
       = \left( 1 + {v^k v_k \over 1 + 2 \varphi} \right)
       {{\cal N}_{,i} \over {\cal N}}
       + {v_{i,k} v^k \over 1 + 2 \varphi}
       - {v^k v_k \varphi_{,i} \over (1 + 2 \varphi)^2}
       + {1 \over a {\cal N}} \sqrt{ 1 + {v^k v_k \over 1 + 2 \varphi} }
       \left[ \left( a v_i \right)_{,0}
       + {v_{i,k} \chi^k \over 1 + 2 \varphi}
       + v_k \left( {\chi^k \over 1 + 2 \varphi} \right)_{,i} \right],
   \label{acceleration} \\
   & & \widetilde \sigma_{ij}
       = - \overline{K}_{ij} \sqrt{ 1 + {v^\ell v_\ell \over 1 + 2 \varphi} }
       + a v_{(i,j)}
       - {a \over 1 + 2 \varphi} \left( v_i \varphi_{,j} + v_j \varphi_{,i}
       - {2 \over 3} v^k \varphi_{,k} \delta_{ij} \right)
       + a v_{(i} \Bigg\{
       \left( 1 + {v^k v_k \over 1 + 2 \varphi} \right)
       {{\cal N}_{,j)} \over {\cal N}}
   \nonumber \\
   & & \qquad
       + {v_{j),k} v^k \over 1 + 2 \varphi}
       - {v^k v_k \varphi_{,j)} \over (1 + 2 \varphi)^2}
       + {1 \over a {\cal N}} \sqrt{ 1 + {v^\ell v_\ell \over 1 + 2 \varphi} }
       \left[ \left( a v_{j)} \right)_{,0}
       + {v_{j),k} \chi^k \over 1 + 2 \varphi}
       + v_k \left( {\chi^k \over 1 + 2 \varphi} \right)_{,j)} \right]
       \Bigg\}
   \nonumber \\
   & & \qquad
       - {1 \over 3} a^2 v_i v_j \left[
       {v^k \varphi_{,k} \over a (1 + 2 \varphi)^2}
       + \left( 3 H - \kappa \right)
       \sqrt{ 1 + {v^k v_k \over 1 + 2 \varphi} } \right]
   \nonumber \\
   & & \qquad
       - {1 \over 3} {a \over {\cal N}}
       \left[ \left( 1 + 2 \varphi \right) \delta_{ij}
       + v_i v_j \right] \left[
       {\left( {\cal N} v^k \right)_{,k} \over 1 + 2 \varphi}
       + \left( \partial_0
       + {\chi^k \over a (1 + 2 \varphi)} \nabla_k \right)
        \sqrt{ 1 + { v^\ell v_\ell \over 1 + 2 \varphi} } \right],
   \label{sigma} \\
   & & \widetilde \omega_{ij}
       = a v_{[i,j]}
       - a v_{[i} \Bigg\{
       \left( 1 + {v^k v_k \over 1 + 2 \varphi} \right)
       {{\cal N}_{,j]} \over {\cal N}}
       + {v_{j],k} v^k \over 1 + 2 \varphi}
       - {v^k v_k \varphi_{,j]} \over (1 + 2 \varphi)^2}
   \nonumber \\
   & & \qquad
       + {1 \over a {\cal N}} \sqrt{ 1 + {v^\ell v_\ell \over 1 + 2 \varphi} }
       \left[ a v_{j],0}
       + {v_{j],k} \chi^k \over 1 + 2 \varphi}
       + v_k \left( {\chi^k \over 1 + 2 \varphi} \right)_{,j]} \right]
       \Bigg\},
   \label{omega}
\eea
where $\partial_0 \equiv {\partial \over \partial \eta}$.

Taking a normal frame with $\widetilde u_a = \widetilde n_a$, we have $v_i = 0$, thus \bea
   & & \widetilde \theta^{(n)} = - K = 3 H - \kappa, \quad
       \widetilde a^{(n)}_i = {{\cal N}_{,i} \over {\cal N}}, \quad
       \widetilde \sigma^{(n)}_{ij} = - \overline{K}_{ij}, \quad
       \widetilde \omega^{(n)}_{ij} = 0.
   \label{kinematic-quantities-normal}
\eea In the normal frame we have to recover the energy flux term $\widetilde q_a$ in equation (\ref{Tab-general}), see below equation (\ref{fluid-n}).


Using the covariant momentum conservation in equation (\ref{eq7-cov}), equations (\ref{acceleration}) and (\ref{omega}) become \bea
   & & \widetilde a_i
       = - {1 \over \widetilde \mu + \widetilde p}
       \left\{
       \widetilde p_{,i}
       + a v_i \left[
       {\widehat \gamma \over {\cal N}} {\partial \over \partial t}
       + {1 \over a ( 1 + 2 \varphi )}
       \left( v^k
       + {\widehat \gamma \over a {\cal N}} \chi^k \right)
       \nabla_k \right] \widetilde p \right\},
   \label{acceleration-2} \\
   & & \widetilde \omega_{ij}
       = a v_{[i,j]}
       + {1 \over \widetilde \mu + \widetilde p}
       a v_{[i} \widetilde p_{,j]}.
   \label{omega-2}
\eea

Parts of the covariant equations are (Ehlers 1961; Hawking 1966;
Ellis 1971, 1973) \bea
   & & \widetilde {\dot {\widetilde \theta}}
       + {1 \over 3} \widetilde \theta^2
       - \widetilde a^a_{\;\; ;a}
       + \widetilde \sigma^{ab} \widetilde \sigma_{ab}
       - \widetilde \omega^{ab} \widetilde \omega_{ab}
       + 4 \pi G \left( \widetilde \mu + 3 \widetilde p \right)
       - \Lambda = 0,
   \label{covariant-Raychaudhury-eq} \\
   & & \widetilde {\dot {\widetilde \mu}}
       + \left( \widetilde \mu + \widetilde p \right)
       \widetilde \theta = 0,
   \label{covariant-E-conserv} \\
   & & \widetilde a_a = - {\widetilde h^b_a \widetilde p_{,b}
       \over \widetilde \mu + \widetilde p}.
   \label{covariant-mom-conserv}
\eea These are the Raychaudhury equation, and the covariant energy conservation and the covariant momentum conservation equations, respectively; the latter two equations are presented in the energy-frame with vanishing energy flux $\widetilde q_a \equiv 0$ (Ellis 1971, 1973; Hwang \& Vishniac 1990); the fluid four vector $\widetilde u_a$ in the energy frame ($\widetilde q_a \equiv 0$) coincides with the normal frame four-vector $\widetilde n_a$ for $v_i = 0$ (the irrotational and the comoving gauge).
By combining equations (\ref{covariant-Raychaudhury-eq})-(\ref{covariant-mom-conserv}) we have (Jackson 1972, 1993; Hwang \& Vishniac 1990)
\bea
   & & {\widetilde {\ddot {\widetilde \mu}}
       \over \widetilde \mu + \widetilde p}
       = 4 \pi G \left( \widetilde \mu + 3 \widetilde p \right)
       - \Lambda
       + {4 \over 3} \widetilde \theta^2
       + \widetilde \sigma^{ab} \widetilde \sigma_{ab}
       - \widetilde \omega^{ab} \widetilde \omega_{ab}
       - \widetilde a^c \widetilde a_c
       + \widetilde h^{ab} \left(
       {\widetilde p_{,a} \over \widetilde \mu + \widetilde p}
       \right)_{;b}.
   \label{Jackson-eq}
\eea Using the covariant kinematic quantities based on $\widetilde
u_a$ presented above we can derive the nonlinear perturbation
equations in gauge-ready forms. Equations (\ref{covariant-E-conserv}) and
(\ref{covariant-mom-conserv}) give equations (\ref{eq6-cov}) and
(\ref{eq7-cov}), respectively. Equation
(\ref{covariant-Raychaudhury-eq}) in general leads to a complicated
combination; in the normal frame it gives equation (\ref{eq4}) using the following relations of the fluid quantities between the energy frame and the normal frame. The fluid quantities in the normal frame are similarly introduced as in equation (\ref{Tab-decomposition}) now based on the normal four-vector as \bea
   & & \widetilde \mu^{(n)} = \widetilde T_{ab} \widetilde n^a
       \widetilde n^b, \quad
       \widetilde p^{(n)} = {1 \over 3} \widetilde T_{ab}
       \widetilde h^{(n)ab}, \quad
       \widetilde q^{(n)}_a = - \widetilde T_{cd}
       \widetilde n^c \widetilde h^{(n)d}_{\;\;\;\;\;a}, \quad
       \widetilde \pi^{(n)}_{ab} = \widetilde T_{cd}
       \widetilde h^{(n)c}_{\;\;\;\;\;a}
       \widetilde h^{(n)d}_{\;\;\;\;\;b} - \widetilde p^{(n)}
       \widetilde h^{(n)}_{ab},
   \label{Tab-decomposition-n}
\eea where $\widetilde h^{(n)}_{ab} \equiv \widetilde g_{ab} + \widetilde n_a \widetilde n_b$. Using equations (\ref{Tab}), (\ref{metric-A}), (\ref{metric-A-inverse}), (\ref{n_a}) and (\ref{u_a}), we have \bea
   & & \widetilde \mu^{(n)} = \widetilde \mu
       + \left( \widetilde \mu + \widetilde p \right)
       {v^k v_k \over 1 + 2 \varphi}, \quad
       \widetilde p^{(n)} = \widetilde p
       + {1 \over 3} \left( \widetilde \mu + \widetilde p \right)
       {v^k v_k \over 1 + 2 \varphi},
   \nonumber \\
   & &
       \widetilde q^{(n)}_i
       = a \left( \widetilde \mu + \widetilde p \right)
       \widehat \gamma v_i, \quad
       \widetilde \pi^{(n)}_{ij}
       = a^2 \left( \widetilde \mu + \widetilde p \right)
       \left( v_i v_j
       - {1 \over 3} v^k v_k \delta_{ij} \right).
   \label{fluid-n}
\eea In the normal frame the energy flux $\widetilde q^{(n)}_i$ plays the role of the fluid velocity ($\widetilde u_i = a v_i$) in the energy frame.

Equations (\ref{eq6-cov}) and (\ref{eq7-cov}) follow from the
covariant energy and momentum conservation equations based on
projecting along $\widetilde u_a$ and $\widetilde h_{ab}$ as \bea
   & & 0 = \widetilde T^b_{a;b} \widetilde u^a
       = - \widetilde {\dot {\widetilde \mu}}
       - \left( \widetilde \mu + \widetilde p \right)
       \widetilde \theta, \quad
       0 = \widetilde T^b_{c;b} \widetilde h^c_a
       = \left( \widetilde \mu + \widetilde p \right) \widetilde a_a + \widetilde h^b_a \widetilde p_{,b}.
\eea Whereas the ADM energy and momentum conservation in equations (\ref{eq6}) and (\ref{eq7}) are based on projecting along $\widetilde n_a$ and $\widetilde h^{(n)}_{ab} \equiv \widetilde g_{ab} + \widetilde n_a \widetilde n_b$. We have the following relations
\bea
   & & 0 = \widetilde T^b_{a;b} \widetilde n^a
       = {1 \over N} \widetilde T^b_{0;b}
       - {N^i \over N} \widetilde T^b_{i;b}
       = - {1 \over N} \left( \widetilde u_0 - N^i \widetilde u_i \right)
       \widetilde T^b_{a;b} \widetilde u^a
       - {1 \over N} \left( {\widetilde u^i \over \widetilde u^0}
       + N^i \right) \widetilde T^b_{a;b} \widetilde h^a_i,
   \nonumber \\
   & & 0 = \widetilde T^b_{a;b} \widetilde h^{(n)a}_{\;\;\;\;\; i}
       = \widetilde T^b_{i;b}
       = \widetilde T^b_{a;b} \widetilde h^a_i
       - \widetilde u_i \widetilde T^b_{a;b} \widetilde u^a.
   \label{conservation-cov-ADM}
\eea These lead to equations (\ref{E-conservation}) and (\ref{Mom-conservation}).

\section{Fluid velocities}
                                \label{sec:velocities}

Here we introduce different definitions of the fluid three-velocity:
$v^i$, $\widehat v^i$ and $\overline v^i$. For relations
among them, see equation (\ref{v-relation}).

(I) Following the convention in the linear perturbation theory, we
have naively introduced $v_i$ as (Bardeen 1988; Hwang \& Noh 2007)
\bea
   & & \widetilde u_i \equiv a v_{i}, \quad
       \widetilde u_0 =
       - a {\cal N} \widehat \gamma
       - {\chi^{k} v_{k} \over 1 + 2 \varphi}, \quad
       \widetilde u^i = {v^{i} \over a (1 + 2 \varphi)}
       + {\chi^{i} \over a^2 {\cal N} (1 + 2 \varphi)}
       \widehat \gamma, \quad
       \widetilde u^0
       = {1 \over a {\cal N}} \sqrt{ 1 + {v^{k} v_{k}
       \over 1 + 2 \varphi} }
       \equiv {1 \over a {\cal N}} \widehat \gamma,
   \label{four-vector}
\eea where $\widehat \gamma$ is the Lorentz factor. The form of the
Lorentz factor shows the nontrivial nature of $v_i$ as the velocity
to the nonlinear order.

(II) The fluid three-velocity $\widehat V^i$ measured by the
Eulerian observer with normal four-velocity $\widetilde n_c$ is
defined as (Banyuls et al 1997) \bea
   & & \widehat V^i
       \equiv {\widetilde h^{(n)i}_{\;\;\;\;\; c} \widetilde u^c
       \over - \widetilde n_c \widetilde u^c}
       = {1 \over N} \left( {\widetilde u^i \over \widetilde u^0} + N^i \right),
\eea where $\widetilde h^{(n)}_{ab} \equiv \widetilde g_{ab} +
\widetilde n_a \widetilde n_b$ is the projection tensor normal to
$\widetilde n^c$, and the index of $\widehat V^i$ is raised and
lowered by the ADM three-space metric $h_{ij}$. In the ADM notation
we have \bea
   & &
       \widetilde u_i = N \widetilde u^0 \widehat V_i, \quad
       \widetilde u_0 = - N \widetilde u^0
       \left( N - N_i \widehat V^i \right), \quad
       \widetilde u^i
       \equiv \widetilde u^0 \left( N \widehat V^i - N^i \right), \quad
       \widetilde u^0 = {1 \over N} {1 \over \sqrt{ 1 - \widehat V^k \widehat V_k}}
       \equiv {1 \over N} \widehat \gamma.
\eea In order to use the perturbation notation, we introduce \bea
   & & \widehat V_i \equiv a \widehat v_i, \quad
       \widehat V^i = {\widehat v^i \over a (1 + 2 \varphi)},
\eea where $\widehat v^i$ is raised and lowered by $\delta_{ij}$.
Thus, we have \bea
   & & \widetilde u_i = a \widehat \gamma \widehat v_{i}, \quad
       \widetilde u_0 =
       - a \widehat \gamma \left( {\cal N}
       + {\chi_k \widehat v^k \over a ( 1 + 2 \varphi )} \right), \quad
       \widetilde u^i = {\widehat \gamma \over a (1 + 2 \varphi)}
       \left( \widehat v^i
       + {1 \over a {\cal N}} \chi^i \right), \quad
       \widetilde u^0
       = {1 \over a {\cal N}} {1 \over \sqrt{ 1 - {\widehat v^k \widehat v_k \over 1 +2 \varphi}}}
       \equiv {1 \over a {\cal N}} \widehat \gamma.
   \label{four-vector-hat}
\eea Using $\widehat v_i$ the Lorentz factor becomes a well known
form.

(III) In the literature, we often find the fluid coordinate three-velocity
introduced as (Wilson 1972; Bardeen 1980; Kodama \& Sasaki 1984)
\bea
   & & \overline V^i \equiv {\widetilde u^i \over \widetilde u^0}
       = {d x^i \over d x^0},
\eea where the index of $\overline V^i$ is raised and lowered by the
ADM three-space metric $h_{ij}$. In the ADM notation we have \bea
   & &
       \widetilde u_i = \widetilde u^0 \left( \overline V_i + N_i \right), \quad
       \widetilde u_0 = - \widetilde u^0 \left[ N^2 - N^k \left( \overline V_k +
       N_k
       \right) \right], \quad
       \widetilde u^i
       \equiv \widetilde u^0 \overline V^i, \quad
       \widetilde u^0 = {1 \over \sqrt{ N^2
       - (\overline V^k + N^k) (\overline V_k + N_k)}}
       \equiv {1 \over N} \widehat \gamma.
\eea In order to use the perturbation notation, we introduce \bea
   & & \overline V^i \equiv \overline v^i, \quad
       \overline V_i = a^2 (1 + 2 \varphi) \overline v_i,
\eea where $\overline v^i$ is raised and lowered by $\delta_{ij}$.
Thus, we have \bea
   & & \widetilde u_i = {a \over {\cal N}} \widehat \gamma
       \left[ \left( 1 + 2 \varphi \right) \overline v_{i}
       - {1 \over a} \chi_i \right], \quad
       \widetilde u_0 =
       - a {\cal N} \widehat \gamma
       \left[ 1 + {1 \over a {\cal N}^2}
       \chi^k \left( \overline v_k - {\chi_k \over a (1 + 2 \varphi)} \right) \right],
   \nonumber \\
   & &
       \widetilde u^i = {\widehat \gamma \over a {\cal N}}
       \overline v^i, \quad
       \widetilde u^0
       = {1 \over a {\cal N} \sqrt{
       1 - {1 + 2 \varphi \over {\cal N}^2}
       \left( \overline v^k - {\chi^k \over a (1 + 2 \varphi)} \right)
       \left( \overline v_k - {\chi_k \over a (1 + 2 \varphi)} \right)}}
       \equiv {1 \over a {\cal N}} \widehat \gamma.
   \label{four-vector-overline}
\eea The $\overline v^i$ is the same as the velocity introduced in our
PN approximation (Hwang et al 2008).

The relations among the velocities $v^i$, $\widehat v^i$ and
$\overline v^i$ are the following \bea
   & & {1 \over \widehat \gamma} v_i
       = \widehat v_i
       = {1 \over {\cal N}} \left[ \left( 1 + 2 \varphi \right) \overline v_i
       - {1 \over a} \chi_i \right].
   \label{v-relation}
\eea The Lorentz factor can be written as \bea
   & & \widehat \gamma
       = \sqrt{ 1 + {v^k v_k \over 1 + 2 \varphi} }
       = {1 \over \sqrt{ 1
       - {\widehat v^k \widehat v_k \over 1 + 2 \varphi}}}
       = {1 \over \sqrt{
       1 - {1 + 2 \varphi \over {\cal N}^2}
       \left( \overline v^k - {\chi^k \over a (1 + 2 \varphi)} \right)
       \left( \overline v_k - {\chi_k \over a (1 + 2 \varphi)} \right)}}.
   \label{Lorentz-factor}
\eea

A three-vector can always be decomposed to a longitudinal and
transverse part as \bea
   & & v_i \equiv - v_{,i} + v^{(v)}_i, \quad
       \widehat v_i \equiv - \widehat v_{,i} + \widehat v^{(v)}_i,
\eea with $v^{(v),i}_i \equiv 0 \equiv \widehat v^{(v),i}_i$, and
similarly for $\overline v_i$. However, the relation in equation
(\ref{v-relation}) shows that $v^{(v)}_i = 0$ does not imply
$\widehat v^{(v)}_i = 0$, and vice versa, and similarly for
$\overline v_i$. Due to the nonlinear relations among different
velocities the scalar- and vector-decompositions for $v_i$ and
$\widehat v_i$ do not coincide with each other to the nonlinear
order. To the nonlinear order the scalar- and vector-type
perturbations are coupled in the equation level.

Our comoving slicing condition was imposed on $v_i$ as $v \equiv 0$.
This corresponds to $\widehat v \equiv 0$ for vanishing vector-type perturbation; in the presence of vector-type perturbation these two conditions differ from each other from the third-order perturbation. However, the
comoving slicing in the variable $\overline v_i$ is more
delicate. Even to the linear order, under our congruence (spatial
gauge) condition, $\overline v$ is not a slicing (temporal gauge)
condition; to the linear order we have $\overline v = v
- \chi/a \equiv v_\chi$ which is already gauge-invariant (under our
congruence condition). To the nonlinear order we should regard $v
\equiv 0$ or $\widehat v \equiv 0$ as the comoving gauge condition,
and the condition in terms of $\overline v$ is rather cumbersome.

To the fully nonlinear order, we have \bea
   & & \widetilde {\dot {\widetilde \mu}}
       = \left[ {1 \over a {\cal N}}
       \widehat \gamma
       \partial_0
       + {1 \over a (1 + 2 \varphi)} \left( v^\ell
       + \widehat \gamma {\chi^\ell \over a {\cal N}}
       \right)
       \nabla_\ell \right]
       \widetilde \mu
       = \widehat \gamma \left[ {1 \over a {\cal N}} \partial_0
       + {1 \over a ( 1 + 2 \varphi )}
       \left( \widehat v^\ell + {\chi^\ell \over a {\cal N}} \right)
       \nabla_\ell \right] \widetilde \mu
       = \widehat \gamma {1 \over a {\cal N}} \left(
       \partial_0
       + \overline v^\ell \nabla_\ell
       \right) \widetilde \mu,
   \nonumber \\
   & & \widetilde h^b_i \widetilde \mu_{,b}
       = \widetilde \mu_{,i}
       + v_i \left[ {\widehat \gamma \over {\cal N}} \partial_0
       + {1 \over 1 + 2 \varphi}
       \left( v^\ell + {\widehat \gamma \over a {\cal N}} \chi^\ell
       \right) \nabla_\ell \right]
       \widetilde \mu
       = \widetilde \mu_{,i}
       + \widehat \gamma^2 \widehat v_i
       \left[ {1 \over {\cal N}} \partial_0
       + {1 \over 1 + 2 \varphi}
       \left( \widehat v^\ell + {\chi^\ell \over a {\cal N}}
       \right) \nabla_\ell \right]
       \widetilde \mu
   \nonumber \\
   & & \qquad
       = \widetilde \mu_{,i}
       + {\widehat \gamma^2 \over {\cal N}^2}
       \left[ \left( 1 + 2 \varphi \right) \overline v_i
       - {1 \over a} \chi_i \right]
       \left( \partial_0
       + \overline v^\ell
       \nabla_\ell \right)
       \widetilde \mu.
   \label{EB-WK-NL}
\eea For comparison, we presented the different expressions based on
three definitions of the fluid velocities.

A covariant spatial gradient variable \bea
   & & \widetilde \Delta_a \equiv {1 \over \widetilde \mu}
       \widetilde h^b_a \widetilde \mu_{,b},
\eea was introduced as a covariant and gauge-invariant variable in
the linear perturbation theory (Ellis \& Bruni 1989; Woszczyna \&
Ku{\l}ak 1989), see also Hawking (1966). Using the gauge
transformation properties to the second order in Hwang et al (2012)
we can show that $\widetilde \Delta_i$ is gauge-invariant {\it only}
to the linear order. To the linear order, we have \bea
   & & \widetilde \Delta_i
       = \nabla_i \delta_v
       + a {\dot \mu \over \mu} v_i^{(v)},
\eea which is a sum of the spatial gradient of $\delta$ in the
comoving gauge and the vector-type perturbation, and is
gauge-invariant; for $\delta_v$, see equation (\ref{GI}). In the
comoving gauge without vector-type perturbation (thus $v_i = 0$) we
have $\widetilde \Delta_{i v} = \delta_{v,i}/(1 + \delta_v)$ which
is related to the density perturbation in the comoving gauge to the
fully nonlinear order.

\section{Multiple components of ideal fluid}
                                    \label{sec:Multiple}

Here we consider the case of multiple components of ideal fluid. Even in the presence of many fluids (with vanishing fluxes and anisotropic stresses) all our equations are valid with the fluid quantities considered as the collective ones. In the presence of $N$ fluids we have
\bea
   & & \widetilde T_{ab} = \sum_J \widetilde T_{(J)ab},
   \label{Tab-Tab-I}
\eea
with the fluid quantities of collective and  individual components introduced as
\bea
   & & \widetilde T_{ab}
       \equiv \widetilde \mu \widetilde u_a \widetilde u_b
       + \widetilde p \left( \widetilde g_{ab}
       + \widetilde u_a \widetilde u_b \right), \quad
       \widetilde T_{(I)ab}
       \equiv \widetilde \mu_{(I)} \widetilde u_{(I)a} \widetilde u_{(I)b}
       + \widetilde p_{(I)} \left( \widetilde g_{ab}
       + \widetilde u_{(I)a} \widetilde u_{(I)b} \right),
\eea
where indices $I, J, \dots = 1, 2 \dots N$ identify the fluid component. From equation (\ref{Tab-Tab-I}) we have
\bea
   & & \widetilde \mu
       = \sum_J \left\{ \widetilde \mu_{(J)}
       + \left( \widetilde \mu_{(J)} + \widetilde p_{(J)} \right)
       \left[ \left( \widetilde u^c_{(J)} \widetilde u_c \right)^2 - 1 \right]
       \right\}, \quad
       \widetilde p
       = \sum_J \left\{ \widetilde p_{(J)}
       + {1 \over 3}
       \left( \widetilde \mu_{(J)} + \widetilde p_{(J)} \right)
       \left[ \left( \widetilde u^c_{(J)} \widetilde u_c \right)^2 - 1 \right]
       \right\},
   \label{fluid-Multi-1} \\
   & & \widetilde \mu - 3 \widetilde p
       = \sum_J \left( \widetilde \mu_{(I)}
       - 3 \widetilde p_{(J)} \right), \quad
       \widetilde u_a
       = - {1 \over \widetilde \mu + \sum_K \widetilde p_{(K)}}
       \sum_J \left( \widetilde \mu_{(J)} + \widetilde p_{(J)} \right)
       \widetilde u^c_{(J)} \widetilde u_c
       \widetilde u_{(J)a}.
   \label{fluid-Multi-u}
\eea
Now we introduce the normalized ($\widetilde u^c_{(I)} \widetilde u_{(I)c} \equiv -1$) fluid four-vector of individual component as
\bea
   & & \widetilde u_{(I)i} \equiv a v_{(I)i}.
   \label{four-vector-I}
\eea The rest of $\widetilde u_{(I)a}$ are the same as in equation
(\ref{u_a}) with $v_i$ replaced by $v_{(I)i}$. Similarly,
for the energy-momentum tensor, the ADM fluid quantities, and the
covariant kinematic quantities of the individual fluid component, we
can replace $\widetilde \mu$, $\widetilde p$ and $\widetilde v_i$ to
$\widetilde \mu_{(I)}$, $\widetilde p_{(I)}$ and $\widetilde
v_{(I)i}$, respectively, in equations
(\ref{Tab-pert})-(\ref{sigma}) with
$\widetilde T_{ab}$, $E$, $\widetilde \theta$, etc., replaced by
$\widetilde T_{(I)ab}$, $E_{(I)}$, $\widetilde \theta_{(I)}$, etc.,
respectively. For the ADM fluid quantities we have \bea
   & & E = \sum_J E_{(J)}, \quad
       J_i = \sum_J J_{(J)i}, \quad
       S = \sum_J S_{(J)}, \quad
       \overline{S}^i_j = \sum_J \overline{S}^{\;\;\;\;\; i}_{(J)j},
\eea but these simple relations do not hold for the covariant
kinematic quantities $\widetilde \theta$, $\widetilde a_i$,
$\widetilde \omega_{ij}$ and $\widetilde \sigma_{ij}$. From equation
(\ref{u_a}) for the collective component and for the
$I$-component, we have \bea
   & & \widetilde u^c_{(I)} \widetilde u_c
       = {v^k_{(I)} v_k \over 1 + 2 \varphi}
       - \sqrt{ \left( 1 + {v^k_{(I)} v_{(I)k} \over 1 + 2 \varphi} \right)
       \left( 1 + {v^\ell v_\ell \over 1 + 2 \varphi} \right)}.
   \label{u-u}
\eea
Equation (\ref{fluid-Multi-u}) gives
\bea
   & & v_i
       = - {1 \over \widetilde \mu + \sum_K \widetilde p_{(K)}}
       \sum_J \left( \widetilde \mu_{(J)} + \widetilde p_{(J)} \right)
       \widetilde u^c_{(J)} \widetilde u_c
       v_{(J)i}.
   \label{fluid-Multi-2}
\eea Equations (\ref{fluid-Multi-1}) and (\ref{fluid-Multi-2}) with
equation (\ref{u-u}) provide the relations between the collective
and individual fluid quantities. Using these relations our fully
nonlinear and exact perturbation equations remain valid even in the
multiple component fluid case.

Now, we have to provide the equations followed by the individual fluid. The energy and the momentum conservation equations for individual component are the ones we need.
The energy and momentum conservation equations follow from $\widetilde T^b_{a;b} = 0$, thus
\bea
   & & \widetilde T^{\;\;\;\; b}_{(I)a;b} \equiv \widetilde I_{(I)a}, \quad
       \sum_J \widetilde I_{(J)a} = 0,
\eea
where $\widetilde I_{(I)a}$ is the interaction terms among fluids. In the ADM and the covariant formulations, the energy and the momentum conservation equations are presented in equations (47) and (48) of Noh \& Hwang (2004), and equations (8) and (9) of Hwang \& Noh (2007), respectively. In ideal fluids, these are \bea
   & & E_{(I),0} N^{-1} - E_{(I),i} N^i N^{-1}
       - K \left( E_{(I)} + {1 \over 3} S_{(I)} \right)
       - \overline{S}^{ij}_{(I)} \overline{K}_{ij}
       + N^{-2} \left( N^2 J^i_{(I)} \right)_{:i}
       = - {1 \over N} \left( \widetilde I_{(I)0}
       - \widetilde I_{(I)i} N^i \right),
   \label{E-conservation-I} \\
   & & J_{(I)i,0} N^{-1} - J_{(I)i:j} N^j N^{-1}
       - J_{(I)j} N^j_{\;\;:i} N^{-1} - K J_{(I)i}
       + E_{(I)} N_{,i} N^{-1}
       + S^{\;\;\;\; j}_{(I)i:j}
       + S^{\;\;\;\; j}_{(I)i} N_{,j} N^{-1}
       = \widetilde I_{(I)i}.
   \label{Mom-conservation-I} \\
   & & \widetilde {\dot {\widetilde \mu}}_{(I)}
       + \left( \widetilde \mu_{(I)} + \widetilde p_{(I)} \right)
       \widetilde \theta_{(I)}
       = - \widetilde u^a_{(I)} \widetilde I_{(I)a},
   \label{covariant-E-conserv-I} \\
   & & \widetilde a_{(I)a}
       + {\widetilde h^{\;\;\;\;\; b}_{(I)a} \widetilde p_{(I),b}
       \over \widetilde \mu_{(I)} + \widetilde p_{(I)}}
       = {\widetilde h^{\;\;\;\;\; b}_{(I)a} \widetilde I_{(I)b}
       \over \widetilde \mu_{(I)} + \widetilde p_{(I)}}.
   \label{covariant-mom-conserv-I}
\eea Compare these with equations (\ref{E-conservation}),
(\ref{Mom-conservation}), (\ref{covariant-E-conserv}), and
(\ref{covariant-mom-conserv}), respectively, for collective ones.
From equations
(\ref{E-conservation-I})-(\ref{covariant-mom-conserv-I}) we can
derive the energy and the momentum conservation equations in
alternative forms. The results are the same as equations
(\ref{eq6}), (\ref{eq7}), (\ref{eq6-cov}) and (\ref{eq7-cov}) with
all fluid quantities $\widetilde \mu$, $\widetilde p$ and $v_i$
replaced by $\widetilde \mu_{(I)}$, $\widetilde p_{(I)}$ and
$v_{(I)i}$, respectively, and add the following contributions from
interactions among fluids to the right-hand-sides of the equations
\bea
   & & - {1 \over a {\cal N}}
       \left( \widetilde I_{(I)0}
       + {\chi^i \over a (1 + 2 \varphi)} I_{(I)i} \right),
   \label{I-1} \\
   & & + I_{(I)i},
   \\
   & & - {1 \over a (1 + 2 \varphi)} v^i_{(I)} I_{(I)i}
       - {1 \over a {\cal N}}
       \sqrt{ 1 + {v^k_{(I)} v_{(I)k} \over 1 + 2 \varphi}}
       \left( \widetilde I_{(I)0}
       + {\chi^i \over a (1 + 2 \varphi)} I_{(I)i} \right),
   \\
   & & + {1 \over \widetilde \mu_{(I)} + \widetilde p_{(I)}}
       \left\{
       I_{(I)i}
       + v_{(I)i} \left[
       {1 \over 1 + 2 \varphi} v^j_{(I)} I_{(I)j}
       + {1 \over {\cal N}}
       \sqrt{ 1 + {v^k_{(I)} v_{(I)k} \over 1 + 2 \varphi}}
       \left( \widetilde I_{(I)0}
       + {\chi^j \over a (1 + 2 \varphi)} I_{(I)j} \right)
       \right]
       \right\},
   \label{I-4}
\eea respectively; we have introduced $I_{(I)i} \equiv \widetilde
I_{(I)i}$ where the spatial index of $I_{(I)i}$ is raised and
lowered by $\delta_{ij}$; $I_{(I)i}$ is the perturbed order quantity
whereas $\widetilde I_{(I)0}$ can include the background order
quantity as $\widetilde I_{(I)0} = I_{(I)0} + \delta I_{(I)0}$.
These equations together with equations (\ref{fluid-Multi-1}) and
(\ref{fluid-Multi-2}) complete the additional equations we need in
the multiple fluid system. The vector variables $v_{(I)i}$ and
$I_{(I)i}$ can be decomposed to the scalar- and vector-type
perturbations as \bea
   & & v_{(I)i} = - v_{(I),i} + v^{(v)}_{(I)i}, \quad
       I_{(I)i} = \delta I_{(I),i} + \delta I^{(v)}_{(I)i},
\eea
with $v^{(v)i}_{(I)\;\;|i} \equiv 0 \equiv \delta I^{(v)i}_{(I)\;\;|i}$.

As in the single component case, to the linear order the scalar-type perturbation $\delta_{(I)}$, $\delta p_{(I)}$, $v_{(I)}$, $\delta I_{(I)0}$ and $\delta I_{(I)}$ depend on the temporal gauge transformation whereas the vector-type perturbations are gauge invariant (Hwang 1991). In addition to the fundamental gauge conditions in equation (\ref{temporal-gauges}), in the multi-component case, we have the following gauge conditions available
\bea
   & & {\rm I\!-\!component\!-\!comoving \; gauge:} \hskip 1.65cm v_{(I)}    \equiv 0,
   \nonumber \\
   & & {\rm uniform\!-\!I\!-\!component\!-\!density \; gauge:} \hskip .6cm \delta_{(I)} \equiv 0,
   \label{temporal-gauges-I}
\eea
to the fully nonlinear order.

\section{Minimally coupled scalar field}
                                               \label{sec:MSF}

Here we present the fully nonlinear and exact formulation in the
case of a minimally coupled scalar field. In section
\ref{sec:MSF-CG} we analysed the case in the comoving gauge.

As the minimally coupled scalar field has no anisotropic stress, the same equations in the fluid formulation in this work are valid even for the minimally coupled scalar field case with the fluid quantities replaced by the ones representing the scalar field. In addition we need the equation of motion in equation (\ref{EOM-MSF}).

The fluid quantities and the equation of motion to the fully nonlinear order can be derived in a gauge-ready form. From Equations (\ref{Tab}) and (\ref{Tab-MSF}) we have \bea
   & & \widetilde u_i
       = - {1 \over \widetilde \mu} \widetilde T_{ib} \widetilde u^b
       = - {\widetilde \phi_{,i} \over \widetilde \phi_{,c} \widetilde u^c}, \quad
       \widetilde h^b_a \widetilde \phi_{,b} = 0, \quad
       0 = \widetilde h^{cd} \widetilde \phi_{,c} \widetilde \phi_{,d}
       = \widetilde \phi^{,c} \widetilde \phi_{,c}
       + {\widetilde {\dot {\widetilde \phi}}}^2.
\eea We set $\widetilde \phi = \phi + \delta \phi$ where $\phi$ is
the background order scalar field. Thus, the comoving gauge ($v
\equiv 0$) together with irrotational condition (thus, $v_i = 0$)
implies the uniform-field gauge ($\delta \phi \equiv 0$) and {\it
vice versa} to the fully nonlinear order; i.e., the uniform-field
gauge implies $v_i = 0$, thus the comoving gauge condition as well
as the $v^{(v)}_i = 0$ condition; the case in the comoving gauge is
analyzed in section \ref{sec:MSF-CG}.

From equations (\ref{Tab-MSF}) and (\ref{Tab-decomposition}) we have
\bea
   & & \widetilde \mu = {1 \over 2} {\widetilde {\dot {\widetilde \phi}}}^2
       + \widetilde V, \quad
       \widetilde p = {1 \over 2} {\widetilde {\dot {\widetilde \phi}}}^2
       - \widetilde V, \quad
       a v_i {\widetilde {\dot {\widetilde \phi}}} = - \widetilde \phi_{,i}, \quad
       \widetilde \pi_{ab} = 0.
   \label{fluids-MSF}
\eea From equations (\ref{v-decomposition}) and (\ref{fluids-MSF}) we have \bea
   & & v = {1 \over a} \Delta^{-1} \nabla^k
       \left( {\widetilde \phi_{,k} / {\widetilde {\dot {\widetilde \phi}}}}
       \right), \quad
       v_i^{(v)}
       = - {1 \over a} {\widetilde \phi_{,i} / {\widetilde {\dot {\widetilde \phi}}}}
       + {1 \over a} \nabla_i \Delta^{-1} \nabla^k
       \left( {\widetilde \phi_{,k} / {\widetilde {\dot {\widetilde \phi}}}}
       \right).
   \label{v}
\eea
It is convenient to have \bea
   & & {\widetilde {\dot {\widetilde \phi}}}
       = \widetilde \phi_{,c} \widetilde u^c
       = {\widetilde \phi_{,i} v^i \over a ( 1 + 2 \varphi )}
       + \widehat \gamma {D \widetilde \phi \over D t}
       = {1 \over \widehat \gamma} {D \widetilde \phi \over D t}, \quad
       a v^k v_k {D \widetilde \phi \over D t}
       = - \widehat \gamma v^i \widetilde \phi_{,i},
   \nonumber \\
   & & \widetilde \phi^{,c} \widetilde \phi_{,c}
       = \widetilde g^{cd} \widetilde \phi_{,c} \widetilde \phi_{d}
       = - {1 \over {\cal N}^2} \dot {\widetilde \phi}^2
       - 2 {\chi^i \over a^2 {\cal N}^2 (1 + 2 \varphi)}
       \dot {\widetilde \phi} \widetilde \phi_{,i}
       + {1 \over a^2 (1 + 2 \varphi)}
       \left( \delta^{ij}
       - {\chi^i \chi^j \over a^2 {\cal N}^2 (1 + 2 \varphi)} \right)
       \widetilde \phi_{,i} \widetilde \phi_{,j},
\eea where we introduced a new derivative
\bea
   & & {D \over Dt}
       \equiv {1 \over {\cal N}}
       \left( {\partial \over \partial t}
       + {\chi^i \over a^2 (1 + 2 \varphi)} \nabla_i \right)
       = {1 \over N} \left( \partial_0
       - N^i \nabla_i \right).
\eea

The equation of motion in equation (\ref{EOM-MSF}) gives \bea
   & & - \widetilde \phi^{;c}_{\;\;\; c}
       = {D^2 \widetilde \phi \over Dt^2}
       + \left( 3 H
       - \kappa \right)
       {D \widetilde \phi \over Dt}
       - {( {\cal N} \sqrt{1 + 2 \varphi}
       \widetilde \phi^{,i} )_{,i} \over a^2 {\cal N} (1 + 2 \varphi)^{3/2}}
       = - \widetilde V_{,\widetilde \phi}
       ( \widetilde \phi ).
   \label{EOM-perturbed}
\eea In order to derive equation (\ref{EOM-perturbed}) it is convenient to use the spacetime connection $\widetilde \Gamma^a_{bc}$ presented in terms of the ADM notations in equation (6) of Noh \& Hwang (2004). We can show that Equation (\ref{eq6}) leads to Equation (\ref{EOM-perturbed}), and equation (\ref{eq7}) is identically satisfied. Equation (\ref{EOM-perturbed}) can be written as \bea
   & & \ddot {\widetilde \phi}
       + \left( 3 H {\cal N}
       - {\cal N} \kappa
       - {\dot {\cal N} \over {\cal N}}
       - {\chi^i {\cal N}_{,i} \over
       a^2 {\cal N} ( 1 + 2 \varphi )} \right) \dot {\widetilde \phi}
       + {2 \chi^i \over
       a^2 ( 1 + 2 \varphi )} \dot {\widetilde \phi}_{,i}
       - {1 \over a^2 (1 + 2 \varphi)}
       \left( {\cal N}^2 \delta^{ij}
       - {\chi^i \chi^j \over a^2 (1 + 2 \varphi)} \right) \widetilde \phi_{,ij}
   \nonumber \\
   & & \qquad
       + \Bigg[ - {{\cal N}^2 \over a^2 (1 + 2 \varphi)}
       \left( {{\cal N}^{,i} \over {\cal N}}
       + {\varphi^{,i} \over 1 + 2 \varphi} \right)
       + \left( 3 H {\cal N}
       - {\cal N} \kappa
       - {\dot {\cal N} \over {\cal N}}
       - {\chi^k {\cal N}_{,k} \over
       a^2 {\cal N} ( 1 + 2 \varphi )} \right)
       {\chi^i \over a^2 (1 + 2 \varphi)}
   \nonumber \\
   & & \qquad
       + \left( {\chi^i \over a^2 (1 + 2 \varphi)}
       \right)^{\displaystyle\cdot}
       + {\chi^k \over a^4 (1 + 2 \varphi)}
       \left( {\chi^i \over 1 + 2 \varphi} \right)_{,k}
       \Bigg] \widetilde \phi_{,i}
       = - {\cal N}^2 \widetilde V_{,\widetilde \phi}
       ( \widetilde \phi ).
   \label{EOM}
\eea

To the background order, we have \bea
   & & \mu = {1 \over 2} \dot \phi^2 + V, \quad
       p = {1 \over 2} \dot \phi^2 - V,
   \label{fluids-BG}
\eea and \bea
   & & \ddot \phi + 3 H \dot \phi + V_{,\phi} = 0.
   \label{EOM-BG-order}
\eea
The entropic perturbation $e$ becomes \bea
   & & e \equiv \delta p - {\dot p \over \dot \mu} \delta \mu
       = \left( 1 + {\ddot \phi \over 3 H \dot \phi} \right)
       \left( {\widetilde {\dot {\widetilde \phi}}}^2 - \dot \phi^2 \right)
       + {2 \ddot \phi \over 3 H \dot \phi}
       \left( \widetilde V - V \right).
\eea

To the third order in perturbations, the complete set of scalar-type perturbation equations of a fluid without anisotropic stress was presented in section \ref{sec:3rd-order} in a gauge-ready form. The same equations are valid even for the minimally coupled scalar field case with the fluid quantities replaced by the ones representing the scalar field. The equation of motion is also contained in the complete set of fluid equation, but it is convenient to present it separately. In the following we provide the fluid quantities and the equation of motion to the third order perturbation in a gauge-ready form. We consider both the scalar- and vector-type perturbations.

To the third order, fluid quantities in equation (\ref{fluids-MSF}) give \bea
   & & v_i = - {1 \over a} {\delta \phi_{,i} \over \dot \phi}
       \Bigg( 1 - {\delta \dot \phi \over \dot \phi}
       + \alpha
       + {\delta \dot \phi^2 \over \dot \phi^2}
       - \alpha {\delta \dot \phi \over \dot \phi}
       - {1 \over a^2 \dot \phi} \chi^i \delta \phi_{,i}
       + {1 \over 2} {1 \over a^2 \dot \phi^2} \delta \phi^{,i} \delta \phi_{,i}
       - {1 \over 2} \alpha^2
       + {1 \over 2 a^2} \chi^i \chi_i \Bigg),
   \label{v-third-order}
\eea and for $\widetilde \mu = \mu + \delta \mu$ and $\widetilde p = p + \delta p$ in equation (\ref{fluids-MSF}), we need \bea
   & & {1 \over 2} {\widetilde {\dot {\widetilde \phi}}}^2
       = {1 \over 2} \dot \phi^2 \left[ 1 - 2 \alpha + 4 \alpha^2
       - \left( v^i v_i + {1 \over a^2} \chi^i \chi_i \right)
       \left( 1 - 2 \alpha - 2 \varphi \right)
       + {2 \over a^2} \chi^i \chi_i \alpha
       - 8 \alpha^3 \right]
   \nonumber \\
   & & \qquad
       + \dot \phi \delta \dot \phi
       \left( 1 - 2 \alpha
       + 4 \alpha^2
       - v^i v_i - {1 \over a^2} \chi^i \chi_i \right)
       + {1 \over 2} \delta \dot \phi^2 \left( 1 - 2 \alpha \right)
       + {1 \over a^2} \dot \phi
       \chi^i \delta \phi_{,i} \left( 1 - 2 \alpha - 2 \varphi
       + {\delta \dot \phi \over \dot \phi} \right),
   \nonumber \\
   & & \widetilde V (\widetilde \phi)
       = V + V_{,\phi} \delta \phi
       + {1 \over 2} V_{,\phi\phi} \delta \phi^2
       + {1 \over 6} V_{,\phi\phi\phi} \delta \phi^3.
   \label{mu-third-order}
\eea The equation of motion in equation (\ref{EOM}) gives \bea
   & & \ddot \phi + 3 H \dot \phi + V_{,\phi}
       + \delta \ddot \phi
       + 3 H \delta \dot \phi
       + \left( V_{,\phi\phi} - {\Delta \over a^2} \right) \delta \phi
       + \left( 3 H \alpha - \dot \alpha - \kappa \right) \dot \phi
       + 2 V_{,\phi} \alpha
   \nonumber \\
   & & \qquad
       = \Bigg\{
       {3 \over 2} H \left[ \alpha^2
       - {1 \over a^2} \chi^i \chi_i \left( 1 - \alpha - 2 \varphi \right)
       - \alpha^3 \right]
       + \left( \alpha
       - {1 \over 2} \alpha^2
       + {1 \over 2 a^2} \chi^i \chi_i \right) \kappa
       - 2 \alpha \dot \alpha
       + {1 \over a^2} \chi^i \left( \dot \chi_i - H \chi_i \right)
       \left( 1 - 2 \alpha - 2 \varphi \right)
   \nonumber \\
   & & \qquad
       - {1 \over a^2} \chi^i \chi_i \left( \dot \alpha + \dot \varphi \right)
       + 4 \alpha^2 \dot \alpha
       + {1 \over a^2} \chi^i \alpha_{,i}
       \left( 1 - 2 \alpha - 2 \varphi \right)
       + {1 \over a^4} \chi^i \chi^j \chi_{i,j}
       \Bigg\} \dot \phi
   \nonumber \\
   & & \qquad
       + \Bigg[ - 3 H \left( \alpha - {1 \over 2} \alpha^2
       + {1 \over 2 a^2} \chi^i \chi_i \right)
       + \kappa \left( 1 + \alpha \right)
       + \dot \alpha \left( 1 - 2 \alpha \right)
       + {1 \over a^2} \chi^i \left( \dot \chi_i
       - H \chi_i + \alpha_{,i} \right) \Bigg] \delta \dot \phi
   \nonumber \\
   & & \qquad
       - {2 \over a^2} \chi^i \delta \dot \phi_{,i} \left( 1 - 2 \varphi \right)
       - {1 \over a^2} \dot \chi^i \delta \phi_{,i} \left( 1 - 2 \varphi \right)
       + {1 \over a^2} \left[
       \left( 2 \alpha - 2\varphi
       - 4 \alpha \varphi
       + 4 \varphi^2
       + {1 \over a^2} \chi^i \chi_i \right) \Delta \delta \phi
       - {1 \over a^2} \chi^i \chi^j \delta \phi_{,ij} \right]
   \nonumber \\
   & & \qquad
       + {1 \over a^2} \left[
       \alpha^{,i} \left( 1 - 2 \varphi \right)
       + \varphi^{,i} \left( 1 + 2 \alpha - 4 \varphi \right)
       \right] \delta \phi_{,i}
       - {1 \over a^2} \left[
       H \left( 1 + 3 \alpha - 2 \varphi \right)
       - \dot \alpha - 2 \dot \varphi - \kappa
       \right] \chi^i \delta \phi_{,i}
   \nonumber \\
   & & \qquad
       - \Bigg[ {1 \over 2} V_{,\phi\phi\phi} \delta \phi^2
       + 2 V_{,\phi\phi} \delta \phi \alpha
       + {1 \over a^2} V_{,\phi} \chi^i \chi_i
       \left( 1 - 2 \varphi \right)
       + {1 \over 6} V_{,\phi\phi\phi\phi} \delta \phi^3
       + V_{,\phi\phi\phi} \delta \phi^2 \alpha
       + {1 \over a^2} V_{,\phi\phi} \chi^i \chi_i \delta \phi \Bigg].
   \label{EOM-third-order}
\eea Equations in section \ref{sec:equations} together with equation (\ref{EOM-third-order}) and the fluid quantities in equations (\ref{fluids-MSF}), (\ref{v-third-order}) and (\ref{mu-third-order}) provide a complete set of third-order perturbation equations in a gauge-ready form.

\end{widetext}

\bsp
\label{lastpage}

\end{document}